\documentclass[
pra,
reprint,
amssymb,
amsfonts,
nolongbibliography,
floatfix
]{revtex4-2}

\usepackage[T1]{fontenc}
\usepackage{mathtools}
\usepackage{graphicx}
\usepackage{color}
\usepackage{bm}
\usepackage{hyperref}
    \hypersetup{
    	bookmarksnumbered,		
    	unicode,			
    	colorlinks,			
    	citecolor=[rgb]{.9,0,.5},	
    	urlcolor=[rgb]{0,0,1},		
    	linkcolor=[rgb]{0,.7,0}		
    	}
\usepackage{physics}
\usepackage[version=4]{mhchem}

\newcommand{\hop}{\mathrm{hop}}
\newcommand{\bz}{\mathrm{BZ}}
\newcommand{\tmop}[1]{\ensuremath{\operatorname{#1}}}
\newcommand{\ee}{\mathrm{e}}
\newcommand{\ii}{\mathrm{i}}
\newcommand{\vac}{\mathrm{vac}}
\newcommand{\mf}{\mathrm{MF}}
\newcommand{\hc}{\mathrm{h.c.}}
\newcommand{\bdg}{\mathrm{BdG}}
\newcommand{\eff}{\mathrm{eff}}

\newcommand{\gu}{\mathrm{Gutzwiller}}
\newcommand{\ct}{\mathrm{const.}}
\newcommand{\vari}{\mathrm{var}}
\newcommand{\phs}{\mathrm{phs}}
\newcommand{\hopp}{\mathrm{h}}
\newcommand{\pair}{\mathrm{p}}
\newcommand{\onsite}{\mathrm{o}}
\newcommand{\loc}{\mathrm{loc}}
\newcommand{\inv}{\mathrm{inv}}

\begin{document}
\title{Topological Higgs Amplitude Modes in Strongly Interacting Superfluids}
\author{Junsen Wang}
\author{Youjin Deng}
\author{Wei Zheng}
\email{zw8796@ustc.edu.cn}
\affiliation{Hefei National Laboratory for Physical Sciences at Microscale and Department of Modern Physics, University of Science and Technology of China, Hefei, Anhui 230026, China}
\affiliation{CAS Center for Excellence and Synergetic Innovation Center in Quantum Information and Quantum Physics, University of Science and Technology of China, Hefei, Anhui 230026, China}
\date{\today}
\begin{abstract}
By studying the 2-dimensional Su-Schrieffer-Heeger-Bose-Hubbard model, we show the existence of topological Higgs amplitude modes in the strongly interacting superfluid phase. Using the slave boson approach, we find that, in the large filling limit, the Higgs excitations and the Goldstone excitations above the ground state are well decoupled, and both of them exhibit nontrivial topology inherited from the underlying noninteracting bands. At ﬁnite fillings, they become coupled at high energies; nevertheless, the topology of these modes are unchanged. Moreover, based on an effective action analysis, we further provide a universal physical picture for the topological character of Higgs and Goldstone modes. Our discovery of the first realization of the topological Higgs mode opens the path to novel investigations in various systems such as superconductors and quantum magnetism.
\end{abstract}
\maketitle

\section{Introduction}
Topological matter has been playing a central role in modern condensed matter physics since its discovery in integer quantum Hall effects \cite{klitzing1980new} four decades ago. In those early days, topological properties manifest themselves via quantized bulk observables, which are directly linked to topological invariants \cite{thouless1982quantized}. Based on Haldane's insight \cite{haldane2008possible} that the central topological phenomenon, i.e., robust edge modes resulting from nontrivial bulk topology, is essentially a \textit{wave effect} not tied to fermions, there is a recent trend to study various systems with no fermionic analog, e.g., topological photonics \cite{lu2014topological,ozawa2019topological}, topological phonons \cite{suesstrunk2015observation,liu2020topological}, topological magnons \cite{shindou2013topological,joshi2019z2,kondo2020nonhermiticity,wang2021topological}, topological mechanics and acoustics \cite{yang2015topological,huber2016topological,ma2019topological}, even topological atmospheric and ocean waves \cite{delplace2017topological}. Particularly, cold atomic systems as quantum simulators \cite{wu2016realization,wang2021realization}, provide a unique possibility to study topological Bose superfluids, whose Bogoliubov excitations in the weak-coupling limit also have a topological band structure \cite{engelhardt2015topological,furukawa2015excitation,di2016topological,pan2016bose,xu2016pi,wu2017bogoliubov,luo2018bosonic,ohashi2020generalized,wang2020pseudo,huang2021interaction,wan2021squeezing,wang2021evidence}. These topological quasi-particles, as bosonic in nature, are similar to topological phonons/magnons, which possess robust edge modes dictated by the bulk-boundary correspondence, and are detectable by spectroscopy measurements. So far, all these studies are focusing on the topology of Nambu-Goldstone modes \cite{nambu1961dynamical,goldstone1961field} in the weak-coupling region, which is gapless at low energies.

One then notices that, spontaneously breaking a continuous symmetry leads to two types of collective excitations: gapless Nambu-Goldstone modes and gapped Higgs modes \cite{weinberg1995quantum}. In the standard model of particle physics, the famous Higgs boson \cite{higgs1964broken}, been elusive for decades, was finally discovered recently \cite{aad2012observation,chatrchyan2012observation}. As its close cousin in condensed matter physics, the Higgs amplitude mode \cite{pekker2015amplitude} also attracts much attention; and have been experimentally found in superconductors \cite{sherman2015higgs,tsuchiya2018hidden,shimano2020higgs}, charge density waves \cite{yusupov2010coherent}, quantum magnets \cite{souliou2017raman,jain2017higgs}, and superfluid $\ce{^3He}$-B phase \cite{avenel1980field,collett2013zeeman}. In the strongly interacting superfluid phase of the Bose-Hubbard model realized in cold atomic systems \cite{fisher1989boson,greiner2002quantum}, the Higgs amplitude mode also has been discussed extensively \cite{altman2002oscillating,huber2007dynamical,pollet2012higgs,liu2015massive,di2018particle}, and its observation has been reported recently using Bragg spectroscopy \cite{bissbort2011detecting} and using lattice modulation \cite{endres2012higgs}.

Can the Higgs amplitude modes also be topologically nontrivial? Here we give an affirmative answer to this question by studying a simple variant of the 2D BHM in the strong-coupling limit, which is easily implemented experimentally in cold-atom platforms. We find that, in the large filling limit, the Higgs modes and the Goldstone modes are well decoupled, and both of them exhibit nontrivial topology inherited
from the background noninteracting bands. At finite fillings, they become coupled at high energies; nevertheless, the topology of these modes are unchanged. Based on an effective action analysis, we further provides a universal physical picture for the topology of Higgs and Goldstone modes, which is also applicable to other symmetry-breaking systems, such as superconductors and quantum magnets.

\begin{figure}[t]
\includegraphics[width=0.4\textwidth]{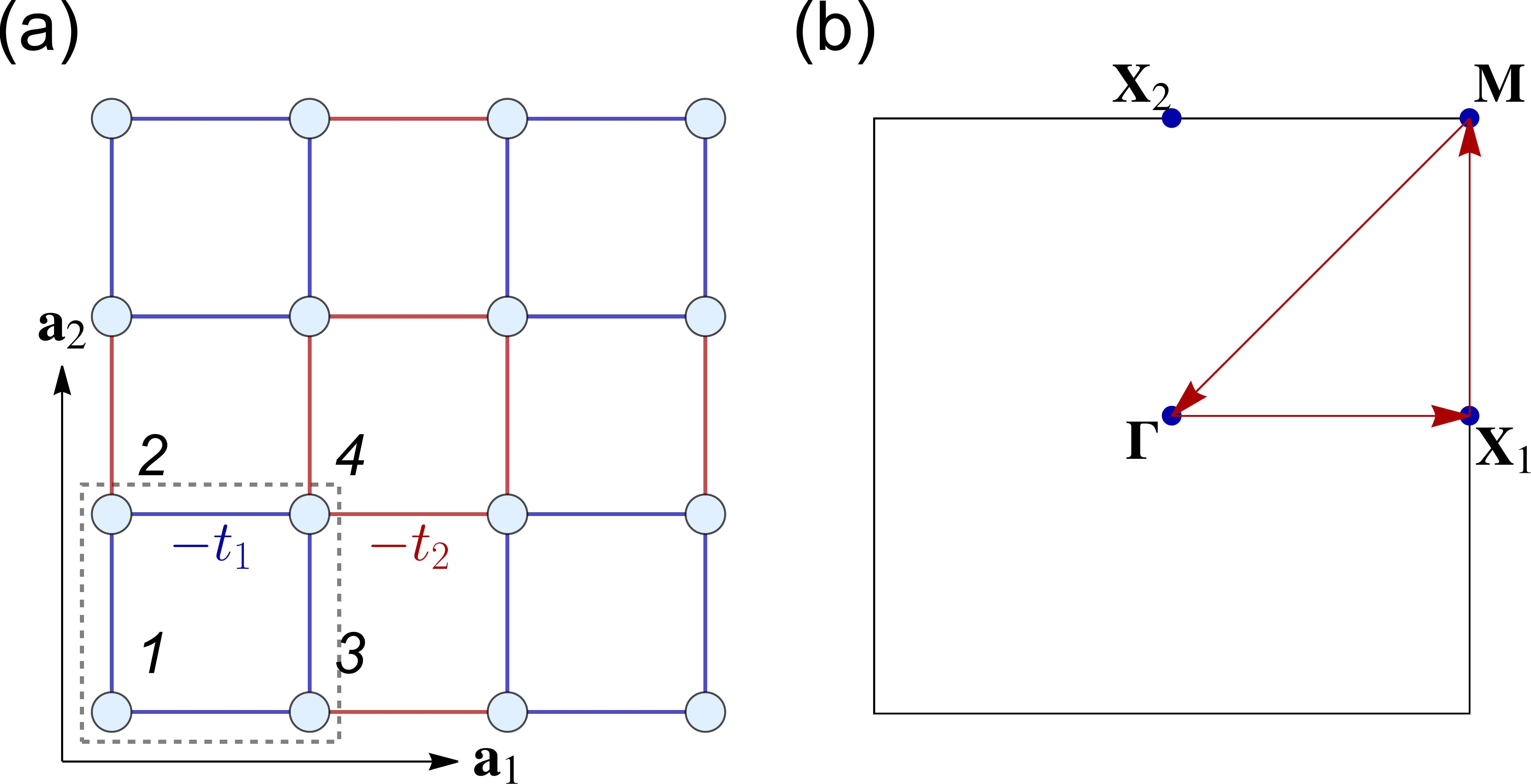}
\caption{(a) 2D SSH-BHM on a square lattice. A unit cell is enclosed by a gray dashed square, with four sublattices labeled by an index $\eta =1,\dots ,4$. The intra-cell (inter-cell) hopping strength is $-t_{1}$ ($-t_{2}$), shown in blue (red) color. Black arrows are two primitive lattice vectors $\vb a_{1,2}$. We set $\abs{\mathbf{a}_{1}}=\abs{\mathbf{a}_{2}}=1$ as the length unit. (b) The first Brillouin zone (BZ) of the model. Four inversion symmetric points are shown explicitly. The red vectors denote the high symmetric path used in Fig.~\ref{fig3}(a,d).}
\label{fig1}
\end{figure}

\section{Model}
As a concrete and minimal example to host topological Higgs amplitude modes, we consider the 2D Su-Schrieffer-Heeger-Bose-Hubbard Model (SSH-BHM), described by the Hamiltonian,
\begin{equation}
\hat{H}=\hat{H}^\hop+\frac{1}{2}U\sum_{i}\hat{n}_{i}(\hat{n}_{i}-1)-\mu \sum_{i}\hat{n}_{i},  \label{Ham}
\end{equation}
where $\hat{H}^{\hop}=-\sum_{ij}t_{ij}\hat{a}_{i}^\dagger\hat{a}_{j}=\sum_{\vb{k}}H_{\vb{k}}^{\hop}\hat{a}_{\vb{k}}^{\dagger}\hat{a}_{\vb{k}}$ is the kinetic term, with the staggered hopping amplitudes along both directions, as depicted in Fig.~\ref{fig1}. This hopping Hamiltonian is the 2D generalization of the SSH model \cite{su1979solitons} introduced in Ref.~\cite{benalcazar2017electric,benalcazar2017quantized,liu2017novel}, whose band topology is protected by the inversion symmetry, $\mathcal{I}H_{-\vb{k}}^{\hop}\mathcal{I}^{-1}=H_{\vb{k}}^{\hop}$, where, in the basis specified by Fig.~\ref{fig1}, the inversion operator reads $\mathcal I = \sigma_1\otimes \sigma_1$ and $\sigma_1$ is the standard Pauli matrix. The corresponding topological invariant is the vectorized Zak phase, also equal to the macroscopic polarization vector \cite{resta1994macroscopic}. Due to inversion symmetry, each component of the polarization vector is restrictedly quantized to a $\mathbb{Z}_{2}$ index \cite{fang2012bulk}. With the additional $\mathrm{C}_4$ symmetry, the polarization center either coincides with the original square lattice (for $t_{1} >t_{2}$, the trivial phase), or coincides with its dual lattice (for $t_{1}<t_{2}$, the topological phase). Note that this topological index can be inferred from the eigenvalues of $\mathcal I$ at inversion symmetric momenta \cite{fu2007topological,fang2012bulk}.

This model Hamiltonian can be realized in experiments by loading spinless bosons in a square optical lattice with the addition of a period-2 superlattice. In the following we will focus on the case where $t_{1}/t_{2}$ is not far from unit. In this region, there is a quantum phase transition between Mott-insulating (MI) and superfluid (SF) phase driven by $t/U$, where $t=(t_{1}+t_{2})/2$, with the superfluid order parameter simply given by $\varphi=\expval{a_i}$ \cite{fisher1989boson}.

\begin{table}[t]
\caption{Parameters used in Eq.~\eqref{hbdg}, for amplitude modes ($\alpha = A$) and phase modes ($\alpha = P$).}
\label{tab1}
\begin{ruledtabular}
\begin{tabular}{cccccccc}
 &$\xi_\alpha$  &$\kappa_\alpha$  &$\zeta_\alpha$ \\
\hline
$\alpha=A$  &  $2z\tilde t \sin^2\theta +\frac{1}{2}U\cos\theta$  &  $\cos^2\theta$  &  $1$  \\
$\alpha=P$  &  $z\tilde t\sin^2\theta + \frac{1}{2}U\cos^2\frac{1}{2}\theta$  &  $\cos^2\frac{1}{2}\theta$  &  $-1$
\end{tabular}
\end{ruledtabular}
\end{table}

\section{Large filling limit}
We utilize the slave boson approach \cite{fresard1994slave,altman2002oscillating,altman2003phase,dickerscheid2003ultracold,huber2007dynamical,pekker2012signatures,huerga2013composite,frerot2016entanglement} to study the excitation spectrums in the SF
phase. The basic idea of the slave boson method is to enlarge the local Hilbert space by introducing bosonic operators, $\hat{b}_{i,n_{i}}^{\dag }$, that create the local Fock state as $\hat{b}_{i,n_{i}}^{\dag }\ket{\vac} = \pqty{\hat{a}_{i}^{\dag }}^{n_{i}}/\sqrt{n_{i}!}\ket{0} $, where $\ket{0} $ is the physical vaccum state and $\ket{\vac} $ is the vaccum state of the slave bosons. The original bosonic operators $\hat{a}_{i}^{\dag }$ then can be expressed in terms of the slave boson operators $\hat{a}_{i}^{\dag }=\sum_{n_{i}}\sqrt{n_{i}+1}\hat{b}_{i,n_{i}+1}^{\dag }\hat{b}_{i,n_{i}}$. To keep the canonical commutation relations of physical bosonic operators, one has to impose a local constraint $\sum_{n_{i}}\hat{b}_{i,n_{i}}^{\dag }\hat{b}_{i,n_{i}}=1 $.

In the vicinity of the SF-MI transition at the $q$th lobe ($q$ is an nonnegative integer), the particle number fluctuation is highly suppressed, such that we can truncate the local Hilbert space by keeping only three relevant states, $\ket{q}_{i}$ and $\ket{q\pm1}_{i}$. Then we consider the large filling limit $q\gg 1$, such that the particle and hole excitations have the same Bose enhancement factors, $\sqrt{q+1}\simeq \sqrt{q}$. Therefore the physical bosons now is given by $\hat{a}_{i}^{\dag }\simeq \sqrt{q}\left( \hat{b}_{i,q+1}^{\dag }\hat{b}_{i,q}+\hat{b}_{i,q}^{\dag }\hat{b}_{i,q-1}\right) $, and the local constrain of the slave bosons becomes $\sum_{\ell =-1}^{1}\hat{b}_{i,q+\ell }^{\dag }\hat{b}_{i,q+\ell }=1$. The slave boson approach starts from the local mean-field Hamiltonian $\hat{H}_{i}^\mf=-zt\varphi \left( \hat{a}_{i}+\hat{a}_{i}^{\dag }\right) +\frac{1}{2}U\hat{n}_{i}\left( \hat{n}_{i}-1\right) -\mu \hat{n}_{i}$. Using slave bosons operators, the mean-field Hamiltonian can be recast as
\begin{equation}
\hat{H}_{i}^\mf= \begin{bmatrix}
    \hat{b}_{i,q+1}^{\dag } &\hat{b}_{i,q}^{\dag} & \hat{b}_{i,q-1}^{\dag }
\end{bmatrix} \begin{bmatrix}
    \frac{U}{2} &  \frac{z\tilde t\varphi}{\sqrt q}  & 0 \\
    & 0 & \frac{z\tilde t\varphi}{\sqrt q}  \\
    \hc &  & \frac{U}{2}
\end{bmatrix} \begin{bmatrix}
    \hat{b}_{i,q+1} \\
\hat{b}_{i,q} \\
\hat{b}_{i,q-1}
\end{bmatrix},
\end{equation}
where $\mu = q-1/2$ is used, corresponding to the so-called particle-hole (PH) symmetric line \footnote{See appendix which includes (1) mean-field theory and phase diagram of the 2D SSH-BHM; (2) derivation of the Bogoliubov-de Gennes Hamiltonian; (3) diagonalization of the BdG Hamiltonian; (4) phase-amplitude character; (5) topological character; (6) a Ginzburg-Landau analysis via strong-Coupling RPA; (7) a brief discussion on validity of slave boson method.}, and a constant term is omitted. It can be self-consistently diagonalized by a rotation
\begin{equation}
    \begin{bmatrix}
        \hat{\beta}_{i,G} \\
        \hat{\beta}_{i,A} \\
        \hat{\beta}_{i,P}
    \end{bmatrix} = \frac{1}{\sqrt{ 2}} \begin{bmatrix}
             \sin \frac{\theta}{2} & \sqrt{2} \cos \frac{\theta}{2} & \sin
               \frac{\theta}{2} \\
             \cos \frac{\theta}{2} & -\sqrt{2}\sin \frac{\theta}{2} & \cos
               \frac{\theta}{2} \\
             1 & 0 & -1
        \end{bmatrix} \begin{bmatrix}
            \hat{b}_{i,q-1} \\
            \hat{b}_{i,q} \\
            \hat{b}_{i,q+1}
        \end{bmatrix},
\end{equation}
where $\theta =\arccos \left( U/16\tilde{t}\right) $ and $\tilde t = q t$. Note the local constraint is preserved under this unitary rotation. One can straightforwardly rewrite the original Hamiltonian Eq.~\eqref{Ham} using these rotated slave bosons. In this representation, the on-site interaction term becomes quadratic, while the hopping term becomes quartic \cite{Note1}. Notice that the rotated slave boson $\hat{\beta}_{i,G}$ generates the mean-field ground state, $\ket{G} =\prod_{i}\hat{\beta}_{i,G}^{\dag }\ket{\vac} $; while $\hat{\beta}_{i,P}$ and $\hat{\beta}_{i,A}$ build up the local excitations. We therefore condense $\hat{\beta}_{i,G}$, namely set $\hat{\beta}_{i,G}\simeq \hat{\beta}_{i,G}^{\dag }\simeq 1$, and treat others as small fluctuations. Then one can expand Eq.~\eqref{Ham} up to quadratic order in the rotated slave bosons, $\hat{H}\simeq \hat{H}^{\left( 2\right) }$, where
\begin{equation}
\hat{H}^{(2)}=\frac{1}{2}\sum_{\vb{k}\in \bz}\begin{bmatrix}
    \hat{\Psi}_{\vb{k},A}^{\dag } & \hat{\Psi}_{\vb{k},P}^{\dag }
\end{bmatrix} H_{\vb{k}}^{\left( 2\right) } \begin{bmatrix}
    \hat{\Psi}_{\vb{k},A} \\
\hat{\Psi}_{\vb{k},P}
\end{bmatrix},  \label{H2}
\end{equation}
where a constant term is omitted and the first order term $\hat H^{(1)}$ vanishes by construction. Here the Nambu spinor is defined by $\hat{\Psi}_{\vb{k},\alpha}=(
    \hat{\beta}_{\vb{k},1,\alpha }, \cdots, \hat{\beta}_{\vb{k},4,\alpha },  \hat{\beta}_{-\vb{k},1,\alpha }^{\dag }, \cdots, \hat{\beta}_{-\vb{k},4,\alpha }^{\dag } ) $, $\alpha=A,P$, and $\hat{\beta}_{\vb{k},\sigma ,\alpha }=\frac{1}{\sqrt{N}}\sum_{l}e^{i\vb k\cdot \vb r_{l}}\hat{\beta}_{l,\eta ,\alpha }$, with $N$ being the total unit-cell number.
In the large filling limit, it turns out that $H_{\vb{k}}^{\left( 2\right) }=H_{\mathbf{k,}A}^{\bdg}\oplus H_{\mathbf{k,}P}^{\bdg}$, i.e., two local excitation modes are decoupled, which enables us to write $\hat{H}^{(2)}=\frac{1}{2}\sum_{\vb{k},\alpha=A,P}\hat{\Psi}_{\vb{k},\alpha }^{\dag }H_{\vb{k},\alpha }^{\bdg}\hat{\Psi}_{\vb{k},\alpha}$. Here $H_{\vb{k},\alpha }^{\bdg}$ takes a particular form,
\begin{equation}
H_{\mathbf{k,}\alpha }^{\bdg}=\begin{bmatrix}
    \xi_{\alpha }+\lambda_{\alpha }H_{\vb{k}}^{\hop} & \zeta_\alpha \kappa_{\alpha
}H_{\vb{k}}^{\hop} \\
\zeta_{\alpha }\kappa_\alpha H_{\vb{k}}^{\hop} & \xi_{\alpha }+\lambda_{\alpha
}H_{\vb{k}}^{\hop}
\end{bmatrix},  \label{hbdg}
\end{equation}
with all parameters given in Table \ref{tab1}. By making the Bogoliubov transformation \cite{Note1}, one can obtain the excitation spectrum above the mean-field ground state, $\hat{H}\simeq \sum_{\mathbf{k,}\lambda
,\alpha =A,P}E_{\mathbf{k,}\lambda ,\alpha }\hat{\gamma}_{\mathbf{k,}\lambda
,\alpha }^{\dag }\hat{\gamma}_{\mathbf{k,}\lambda ,\alpha }$, where $\lambda
$ is the band index, and a constant term is omitted.

\begin{figure}[h!]
\includegraphics[width=0.4\textwidth]{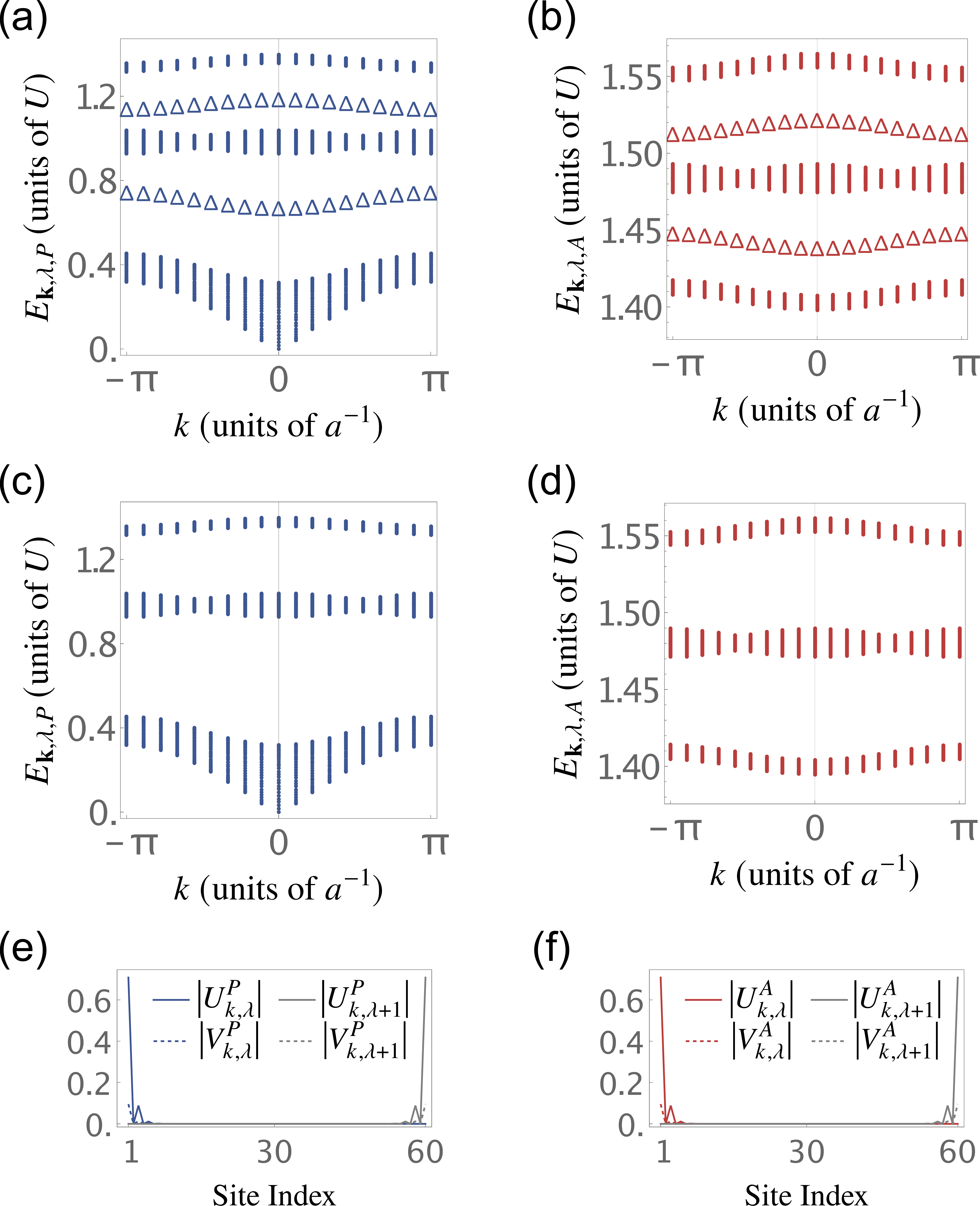}
\caption{Excitation spectrum of the 2D SSH-BHM in a ribbon geometry in the large filling limit for $\tilde t = 3U/16$. (a) ($t_1/t_2=1/8$) and (c) ($t_1/t_2=8$) are the Goldstone phase modes. (b) ($t_1/t_2=1/8$) and (d) ($t_1/t_2=8$) are the Higgs amplitude modes. Bulk modes are shown in solid dots, while topological edge modes are shown in triangles. All modes are doubly degenerate due to inversion symmetry. (e) and (f) are typical wavefunctions of topological edge modes for phase modes and amplitude modes, respectively. And the gray lines are their inversion symmetric partners.}
\label{fig2}
\end{figure}

To identify these excitation modes, $\hat{\gamma}_{\vb{k,}\lambda,\alpha }$, to be amplitude modes (Higgs modes) or phase modes (Goldstone modes), we study the time evolution after a small perturbation above the ground state, $\ket{ \Psi_{\vb{k,}\lambda ,\alpha }( t)} =\ee^{-\ii\hat{H}t}\pqty{ \ket{G} +\epsilon \hat{\gamma}_{\vb{k},\lambda ,\alpha}^{\dag }\ket{G} }$, with $\epsilon\ll 1$. In particular, fluctuation of the order parameters, $\delta \varphi (t)= \expval{\hat a_i}{\Psi _{\mathbf{k,}\lambda ,\alpha }\left( t\right)} -\expval{\hat a_i}{G}$, to the leading order in $\epsilon $, is found to be
\begin{equation}
\delta \varphi \left( t\right) \propto \delta \varphi_{\mathrm{R}}\cos\left( E_{\mathbf{k,}\lambda ,\alpha }t\right) +\delta \varphi _{\mathrm{I}}\sin \left( E_{\mathbf{k,}\lambda ,\alpha }t\right) \ii 
\end{equation}
where $\delta \varphi_{\mathrm{R,I}}=\expval{\hat{\gamma}_{\vb{k},\lambda ,\alpha }\hat{a}_{i}\pm \hat{a}_{i}\hat{\gamma}_{\vb{k},\lambda ,\alpha }^{\dag }}{G}$ \cite{Note1}. Without loss of generality, we can choose the order parameter to be real. Then, if $\delta \varphi \left( t\right) $ is real, the excitation is a pure amplitude mode; while if $\delta \varphi \left( t\right) $ is purely imaginary, the excitation is a pure phase mode. In general, the excitation could be a mixing of both, such that the order parameter fluctuation $\delta \varphi \left( t\right) $ is a generic c-number. Therefore we define a flatness parameter,
\begin{equation}\label{flatness}
F=\frac{\abs{\delta \varphi _{\mathrm{R}}} - \abs{\delta \varphi _{\mathrm{I}}} }{\abs{\delta \varphi _{\mathrm{R}}} + \abs{\delta \varphi _{\mathrm{I}}} }\in [ -1,1],
\end{equation}
to quantify the amplitude and phase components of an excitation. A positive (negative) flatness indicates dominant amplitude (phase) character. A pure amplitude (phase) oscillation corresponds to $F=1\left( -1\right) $. In the large filling limit, by calculating the flatness explicitly, we find $\hat{\gamma}_{\mathbf{k,}\lambda ,P}$ is a pure phase mode, and $\hat{\gamma}_{\mathbf{k,}\lambda ,A}$ is a pure amplitude mode \cite{Note1}.

Excitations above ground state are described by the quadratic Hamiltonian Eq.~\eqref{H2}, thus their topological character is obtained by analyzing the BdG matrix Eq.~\eqref{hbdg}, which also enjoys the inversion symmetry $\mathcal I_\tau H_{-\vb{k},\alpha }^{\bdg}\mathcal I_\tau ^{-1}=H_{\vb{k},\alpha }^{\bdg}$, where $\mathcal I_\tau =\tau_{0}\otimes \mathcal{I}$, and $\tau_0$ is the 2-by-2 identity matrix acting on the Nambu space. We naturally generalize the polarization vector to a symplectic form \cite{engelhardt2015topological}, $\vb P=\int_{\bz}\frac{\dd[2]{k}}{(2\pi)^2 }\vb A(\vb k)$ where $\vb A(\vb{k})=\ii \sum_{\lambda _{1}\leq \lambda \leq \lambda_{2}}\tmop{Tr}\left( {\Gamma _{\lambda }W_{\vb k}^{-1}\partial_{\vb k} W_{\vb{k}}}\right) $, $\Gamma_\lambda$ projects to the $\lambda$th band, and the pseudo-unitary matrix ${W_{\vb{k}}}$ diagonalizes $H_{\vb{k},\alpha }^{\bdg}$. Each component of $\mathbf{P}$ is restrictedly quantized to a $\mathbb{Z}_{2}$ number \cite{Note1}, $P_{\mu }=\frac{1}{2}\bqty{\pqty{{\sum_{\lambda_{1}\leq \lambda\leq \lambda_{2}}n_{\lambda ,\mu }}} \mod 2}$, where $(-1)^{n_{\lambda ,\mu }}=\eta_{\lambda }(\vb X_{\mu })\eta_{\lambda }(\vb \Gamma )$ and $\eta$ is the eigenvalue of the generalized inversion operator $\mathcal I_\tau $. In the large filling limit, one can explicitly polar decompose the Bogoliubov transformation matrix $W_{\vb{k},\alpha}$ as the product of a unitary matrix and a Hermitian (also pseudo-unitary) matrix. It then follows straightforwardly that \cite{Note1}
\begin{equation*}
{W_{\vb{k}_{\inv }}^{-1}}\mathcal I_\tau {W_{\vb{k}_{\inv }}=\tau}_{0}{\otimes }\left( {Q}_{\vb{k}_{\inv }}^{-1 }\mathcal{I}{Q}_{\vb{k}_{\inv }}\right),
\end{equation*}
where $Q_{\vb k}$ is the unitary matrix that diagonalize $H^\hop_{\vb k}$. Namely, the parity eigenvalues of both the Higgs bands and the Goldstone bands at inversion symmetric momenta are identical to the noninteracting bands. Consequently, not only the Goldstone bands, but also the Higgs bands inherit the topology of the background non-interacting bands: When $t_{1}<t_{2}$ ($t_1>t_2$), both of them are topologically non-trivial (trivial), with symplectic polarization vector $\mathbf{P}=\left( 1/2,1/2\right)$ [$\mathbf{P}=\left( 0,0 \right)$]. We confirm the bulk-boundary correspondence numerically by calculating the excitation spectrum in a ribbon geometry and indeed observe the edge states in the topologically non-trivial regime, as shown in Fig.~\ref{fig2}. We also numerically verify that the flatness of the Goldstone bands and the Higgs bands is indeed $+1$ and $-1$, respectively, reflecting the fact that they are fully decouple in the large filling limit.

\begin{figure*}[t]
\includegraphics[width=0.9\textwidth]{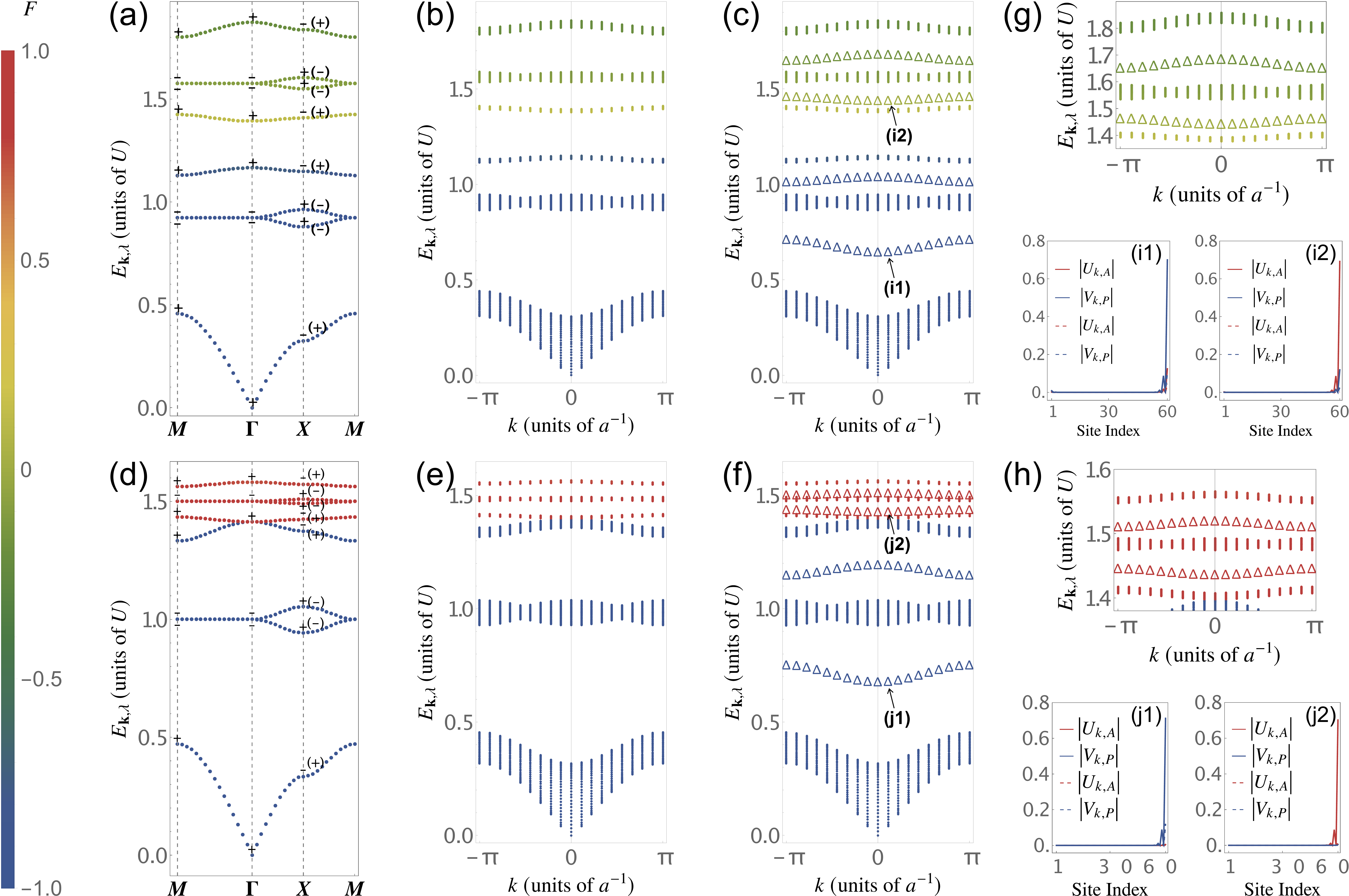}
\caption{Excitation spectrum of the 2D SSH-BHM under PBCs (a,d) and in
a ribbon geometry (b,c,e,f) at filling $q=1$ for (a-c) and $q=50$ for (d-f), for $\tilde t = 3U/16$, with color code indicating the flatness defined in Eq.~\eqref{flatness}. Bulk (topological edge) modes are plotted in dots (triangles), with typical wave functions shown in (i) and (j) (only one of the two degenerate modes is plotted). In (g) and (h) we room in excitation spectrums at high energies for (c) and (f), respectively.}
\label{fig3}
\end{figure*}

\section{Finite filling case}
The slave boson method also works at finite fillings, but now $\hat{a}_{i}^{\dag }\simeq \left( \sqrt{q+1}\hat{b}_{i,q+1}^{\dag }\hat{b}_{i,q}+\sqrt{q}\hat{b}_{i,q}^{\dag }\hat{b}_{i,q-1}\right) $, and the PH symmetric line is given by $\mu = (q-1/2) -\bqty{z\tilde t+ \pqty{\sqrt{q+1}+\sqrt{q}}^{-2} }/4 $ \cite{Note1}, which is bended downwards due to asymmetric Bose enhancement. The resulting excitation spectrum for open (OBCs) and periodic (PBCs) boundary conditions are given in Fig.~\ref{fig3}, which shows that the flatness is between $-1$ and $+1$ in general, due to off-diagonal coupling terms between Goldstone mode and the Higgs mode. When we increasing the filling, this coupling become weaker, such that the flatness tends to $\pm 1$. Despite the absence of pure phase-amplitude character, we can still identify topological character, since $H_{\vb{k}}^{\left( 2\right) }$ in Eq.~\eqref{H2} enjoys an inversion symmetry at any fillings, and the bulk topological index is well defined. By examining the parity eigenvalues at the inversion symmetric momenta, labeled in Fig.~\ref{fig3}(a,d), we find that these excitation bands, either phase or amplitude fluctuation dominated, all inherit the topology of the background noninteracting bands, and they has the same topologically trivial-nontrivial transition point as the background bands. It follows that two groups of mid-gap edge states under OBCs are observed in Fig.~\ref{fig3}(c,f). Their flatness respectively approaches $\pm 1$ when increasing the filling. Thus the coupling between the Higgs bands and the Goldstone bands will not break the topology of excitation spectrum and the bulk-boundary correspondence.

\section{Effective action analysis}
Lastly we present a simple and unified picture for the results obtained so far. At an integer filling $q$, near the SF-MI phase transition, one can use a strong-coupling random-phase approximation \cite{sengupta2005mott} to arrive at an effective action for the 2D SSH-BHM,
\begin{align}
    S_\eff &= \int \dd{\tau}\bigg\{ \sum_{i} \bqty{a_{i}^{\ast }\pqty{\sum_{\ell =0}^{\infty }(-1)^{\ell+1 }c_{\ell }\partial _{\tau }^{\ell }} a_{i} +\frac{\tilde{U}}{2}\abs{a_i}^{4}}\nonumber\\
    &\quad -\sum_{ij}t_{ij}a_{i}^{\ast }a_{j}\bigg\},  \label{Seff}
\end{align}
where all $c_{\ell }$ are real, and $\tilde U$ is a renormalized interaction strength \cite{Note1}. Note this effective action is obtained by \textit{two} successive Hubbard-Stratonovich transformation, and the auxiliary field $a$ in Eq.~\eqref{Seff} generates same correlators as the original bosonic field. In the SF phase, by introducing small fluctuations, $a_{i}(\tau) =( \varphi +\delta \rho
_{i}(\tau) ) \ee^{\ii\delta \theta_{i}( t) }$, we expand Eq.~\eqref{Seff} to quadratic order,
\begin{widetext}
\begin{equation}
S^{(2)}=\int \dd{\tau} \sum_{ij}\begin{bmatrix}
    \delta\rho _{i} & \delta\theta _{i}
\end{bmatrix} \begin{bmatrix}
    t_{ij}-\sum_{n=0}^{\infty }c_{2n}\partial_{\tau }^{2n}-\tilde{\mu}+3\tilde{U}\varphi^{2} & \ii\varphi \delta_{ij}\sum_{n=0}^{\infty }c_{2n+1}\partial_{\tau }^{2n+1} \\
    -\ii\varphi \delta _{ij}\sum_{n=0}^{\infty }c_{2n+1}\partial_{\tau}^{2n+1} & t_{ij}+ \pqty{-\sum_{n=0}^{\infty }c_{2n}\partial_{\tau }^{2n}-\tilde{\mu}+\tilde{U}\varphi^{2}}\varphi^{2}
\end{bmatrix} \begin{bmatrix}
    \delta\rho_{j} \\
    \delta\theta_{j}
\end{bmatrix}.  \label{hij}
\end{equation}
\end{widetext}
By requiring $c_{1}=0$ (i.e., on the PH symmetric line) and in the low-energy limit $\omega \rightarrow 0$ (i.e., dropping all the higher-order time derivative terms), phase mode and amplitude mode becomes decoupled, which explains the persistence of pure phase modes at low energies in Fig.~\ref{fig3} for both fillings. Noticing that along the PH symmetric line, it is easy to show that $c_{\ell }\sim \mathcal{O}(q^{-\ell +1})$ for $\ell \geq 2$, thus in the large filling limit only $c_{2}$ survives, even away from the low-energy limit. Moreover, these modes inherit band topology directly from the 2D SSH model, as the hopping terms not altered. This explains the existence of topological Higgs-amplitude and Goldstone-phase bands for $q\rightarrow \infty $, in Fig.~\ref{fig2}.

As a by-product, we point out that if the hopping term breaks time-reversal symmetry (TRS), i.e., $t_{ij}$ is not purely real, one can easily show that there are off-diagonal terms entering Eq.~\eqref{hij}, which is proportional to the imaginary part of $t_{ij}$. Thus the amplitude modes and phase modes have non-vanishing coupling \emph{even in the infinite filling limit}. This fact can also be derived from the slave boson picture \cite{Note1}. Physically speaking, the so-called PHS indicates that the action Eq.~\eqref{Seff} is invariant under the exchange $\psi \leftrightarrow \psi^{\ast}$, up to a total time derivative term. If $t_{ij} = t_{ji}^* \neq t_{ji}$, the hopping term can not return to itself upon this exchange: $a_i^* t_{ij} a_j \rightarrow a_i t_{ij} a_j^* = a_i^* t_{ji} a_j \neq a_i^* t_{ij} a_j $, where Einstein summation rule is assume. Nevertheless, we predict that the mixed excitation spectrums still inherit the band topology, and will exhibit two groups of mid-gap edge states under OBCs.

\section{Discussion and outlook}
Since the edge modes of Higgs type and Goldstone type illustrated in this work is of topological origin, we expect that they are robust against disorder \cite{peano2018topological} that (1) respect the inversion symmetry (2) and is sufficiently weak so that topological excitation band gap does not close and the system does not enter into other possible phases such as the Bose glass phase (where the topology of excitations may change dramatically). Ultimately, the fate and robustness of topological Higgs modes subject to disorder need a separate, series study in the future. We note that similar works on other bosonic topological system with disorder has been discussed recently \cite{akagi2020topological,wang2020bosonic}.

Thanks to the fast development of experimental techniques,
the Bragg spectroscopy \cite{bissbort2011detecting} and the lattice-modulation spectroscopy \cite{endres2012higgs} can detect the Higgs mode; and the box trap with sharp boundary has been achieved in cold atom systems \cite{gaunt2013bose}. We expect that the predicted topological Higgs amplitude edge modes can be observed as a sharp peak within the band gap in the spectroscopy. In this paper, our discussion are limited to the quadratic order, so that there
is no coupling between these modes. By including higher order terms, interactions between the
excitation modes can be considered. Then
it is interesting to investigate the impacts of mode coupling on the
stability of the highly localized Higgs and Goldstone edge excitations. It is
also interesting to explore similar topological Higgs amplitude modes in
other symmetry breaking systems, such as superconductors and quantum magnets; in particular, for the latter, it is possible that the topological Higgs amplitude modes yield a nontrivial contribution to the thermal Hall effect, similarly to those resulting from the topological magnon \cite{katsura2010theory}.

\begin{acknowledgments}
	We thank Wei Yi, Jinyi Zhang and Zhe-Yu Shi for
helpful comments. YD acknowledges the support by National Natural Science
Foundation of China (under Grant No. 11625522) and the National Key R\&D
Program of China (under Grants No. 2016YFA0301604 and No. 2018YFA0306501).
\end{acknowledgments}


\appendix
\section{Mean-field theory and phase diagram}
In this section, we discuss the mean-field phase diagram of the $d$-dimensional SSH-BHM, which returns to the well-known phase diagram of the standard BHM upon setting $t= (t_1+t_2)/2$ (for $t_1/t_2$ not far from unit).

In the strong-coupling mean-field theory, one decouples the hopping term $-t_{ij} \hat a_i^\dagger \hat a_j$ as $-t_{ij} (\hat a_i^\dagger \phi_j + \hat a_j \phi_i^* -\phi_i^*\phi_j)$. Then the original Hamiltonian Eq.~\eqref{Ham} given in the main text becomes
\begin{eqnarray}\label{hmf}
    \hat H \approx \sum_i \hat H_i^\mf &= &\sum_i\bigg[-\sum_j\pqty{\hat a_i^\dagger t_{ij} \phi_j - \frac{1}{2}t_{ij} \phi_i^*\phi_j + \hc}\nonumber \\
    &&+\frac{1}{2}U \hat n_i (\hat n_i -1) - \mu  \hat n_i\bigg],
\end{eqnarray}
where $\phi_i = \prescript{}{i}{\bra{\Phi_0}}\hat a_i \ket{\Phi_0}_i $ and $\ket{\Phi_0}_i$ is the ground state of $\hat H_i^\mf$ obtained self-consistently. Note this approach is equivalent to introducing the Gutzwiller ansatz $\ket{\Psi_\gu} =  \otimes_i \pqty{\sum_n c_{i,n}\ket{n}_i}$, where $\ket{n}_i=(\hat a^\dagger_i)^n\sqrt{n!}\ket{0}$ ($\ket{0}$ is the vacuum of operator $\hat a$), and minimizing the variational ground state energy $\bra{\Psi_\gu}\hat H\ket{\Psi_\gu}$. Also note Eq.~\eqref{hmf} can be used for the system under open boundary conditions (OBCs), in which case the order parameter $\phi_i$ is generally site-dependent.

For the $d$-dimensional SSH-BHM, the hopping matrix $t_{ij}$ is chosen staggered as $t_1$ and $t_2$ along all $d$ directions. Under periodic boundary conditions (PBCs) and assuming a site-independent, real order parameter $\phi_i = \phi\in\mathbb R$, the mean-field Hamiltonian Eq.~\eqref{hmf} then reduces to
\begin{equation}\label{hmf1}
    \hat H_i^\mf \approx \sum_i\bqty{-\pqty{zt\phi \hat a_j+\hc} + zt \phi^2 +\frac{1}{2}U \hat n_i (\hat n_i -1) - \mu  \hat n_i},
\end{equation}
where $z=2d$ is the coordination number and $t=\frac{t_1+t_2}{2}$. Equation \eqref{hmf1} is precisely the strong-coupling mean-field Hamiltonian for the $d$-dimensional BH model \cite{van2001quantum}, whose phase diagram is reviewed below.

\begin{figure}[t]
    \includegraphics[width=.4\textwidth]{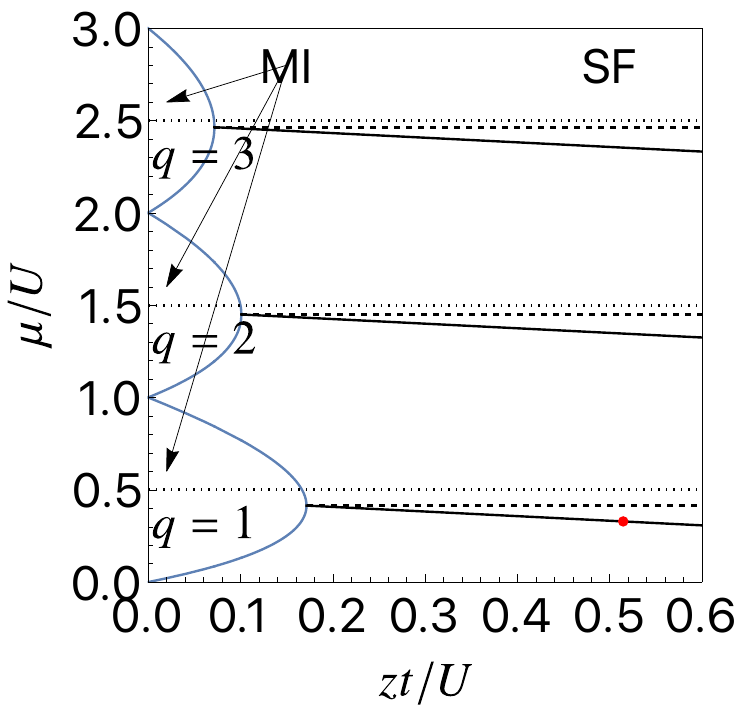}
    \caption{Phase diagram of the $d$D SSH-BHM, with blue lines separating the MI phase and the SF phase. It is the same as the standard BH model with the hopping parameter $t=(t_1+t_2)/2$. Here the coordination number $z=2d=4$ for the 2D SSH-BHM. The black dotted horizontal lines start from each middle of the lobe. The black dashed horizontal lines start from each tip of the lobe. The black solid lines are integer filling lines in the SF phase obtained from Eq.~\eqref{phsmu}. We always set the chemical potential $\mu$ on these black solid lines, where the excitations have the most visible phase-amplitude character. In particular, the red dot corresponds to the case in Fig~3(a-c) the main text.}
    \label{bhmphasediagram}
\end{figure}

Assuming the quantum phase transition being of second order, i.e., $\phi\ll 1$ near the transition boundary, one can treat $\hat V = -zt\phi \hat a_j +\hc$ as the perturbation to $\hat H_0 = zt \phi^2 +\frac{1}{2}U \hat n_i (\hat n_i -1) - \mu  \hat n_i$ in Eq.~\eqref{hmf1}. The unperturbed ground state energy is then given by $\varepsilon_0^{(0)} = \frac{1}{2}Uq(q-1)-\mu q $ for $q=\lfloor \mu/U \rfloor +1$ if $\mu>0$ and $q=0$ otherwise, with the unperturbed ground state being $|\Phi_0^{(0)}\rangle=\ket{q} $. The first order correction vanishes by inspection. While the second order correction is given by the standard formula \cite{landau2013quantum} $\varepsilon_0^{(2)} = \sum_{l\neq 0} \frac{|\langle\varepsilon_0^{(0)}|\hat V|\epsilon_l^{(0)}\rangle|^2}{\varepsilon_0^{(0)} -\epsilon_l^{(0)} }$, which leads to $\varepsilon_0^{(2)} = (zt\phi)^2 \bqty{\frac{q}{(q-1)-\mu}+\frac{q+1}{\mu-qU}}$. Thus the ground state energy reads $ \varepsilon_0 = a_0 + a_2 \phi^2 + \mathcal O(\phi^4)$. According to Landau theory, phase transition occurs at $a_2= (zt)^2 \bqty{\frac{q}{(q-1)-\mu}+\frac{q+1}{\mu-qU}}+zt=0 $, whose solution gives the well-known lobe in the $\mu-zt$ phase diagram at filling $q$ (we set $U=1$ as the energy unit) \cite{fisher1989boson},
\begin{equation}\label{pb}
    \mu_\pm(q) = q-\frac{1}{2} -\frac{1}{2}zt\pm \frac{1}{2}\sqrt{(zt)^2+1-2zt(2q+1)}.
\end{equation}
The \emph{tip} of the lobe corresponds to $\mu_+(q) = \mu_-(q)$, which leads to
\begin{equation}
    (zt_c,\mu_c) = \big[1+2q-2\sqrt{q(1+q)},\sqrt{q(1+q)}-1\big] .
\end{equation}
In the superfluid (SF) phase, we can numerically obtain the line of integer filling factor near the $q$th lobe by setting $ \prescript{}{i}{\bra{\Phi_0}}\hat n_i \ket{\Phi_0}_i =q$. It generally bends downward due to particle-hole (PH) asymmetry at a finite filling. When truncated to only three local states, it has an analytical expression given in Eq.~\eqref{phsmu}. A typical phase diagram is shown in Fig.~\ref{bhmphasediagram}.

\section{Derivation of the Bogoliubov-de Gennes Hamiltonian}
In this section, we derive the bosonic Bogoliubov-de Gennes (BdG) Hamiltonian used in the main text. PBCs and a site-independent order parameter are assumed throughout; differences occurred in OBCs are mentioned in the end.
 
Focusing on the strongly coupled SF phase in the vicinity of the $q$th Mott lobe, only three local states
\begin{equation}\label{trunc3}
    \ket{q-1}_i,\quad\ket{q}_i,\quad\ket{q+1}_i,
\end{equation}
at each site $i$ dominate the low-energy behavior of the system. Following Altman and Auerbach \cite{altman2002oscillating}, we truncate the bosonic Fock space to these three states, and introduce a Gutzwiller-type mean-field ground state ansatz $\ket{G} = \otimes_i \ket{\Phi_0}_i $, where \cite{huber2007dynamical}
\begin{eqnarray}\label{phi0}
    \ket{\Phi_0}_i &=& \cos(\theta/2) \ket{q} + \sin(\theta/2) \big[\cos(\chi/2)\ket{q+1}_i\nonumber \\
    && + \sin(\chi/2) \ket{q-1}_i\big].
\end{eqnarray}
Then the order parameter becomes
\begin{equation}\label{op}
    \phi = \prescript{}{i}{\bra{\Phi_0}}\hat a_i \ket{\Phi_0}_i = \frac{1}{2} \pqty{\sqrt{q+1}\cos\frac{\chi}{2}+\sqrt{q}\sin\frac{\chi}{2}}\sin\theta,
\end{equation}
and the variational ground state energy per site is
\begin{eqnarray}\label{gse}
    \varepsilon_\vari(\theta,\chi) &=& \frac{\expval{\hat H}{G}}{N}\nonumber\\
    & =& \bqty{\frac{1}{2}- \delta\mu \cos\chi} \sin^2\frac{\theta}{2} -\frac{z\tilde t}{2}\bigg[1+\sqrt{1+q^{-1}}\sin\chi \nonumber\\
    && +\frac{1}{2q}(1+\cos\chi)\bigg] \sin^2\theta + \ct,
\end{eqnarray}
where $N$ is the number of lattice sites, two parameters $\tilde t = q t$ and $\delta \mu = \mu - (q-1/2)$ are the renormalized hopping strength and the chemical potential measured from the \emph{middle} of the lobe, respectively. Minimizing Eq.~\eqref{gse} with respect to $\chi$ at a fixed $\theta$, namely, setting $\partial_\chi \varepsilon_\vari(\theta,\chi) =0$, one obtains
\begin{equation}\label{chitheta}
    \chi(\theta) = \arctan\bqty{\frac{2z\tilde t(1+\cos\theta)\sqrt{q(q+1)}}{z\tilde t+\frac{4}{q}\delta\mu +z\tilde t\cos\theta}}.
\end{equation}
Further setting $\partial_\theta \varepsilon_\vari(\theta,\chi)=0$ and using Eq.~\eqref{chitheta}, one can find the mean-field solution $\bar \theta$, whose explicit expression is lengthy and omitted. Note by expanding $\partial_\theta \varepsilon_\vari(\theta,\chi(\theta))$ around $\theta=0$ as $\partial_\theta \varepsilon_\vari(\theta,\chi(\theta)) = \ct + \tilde a_2 \theta +\mathcal O(\theta^2)$, where
\begin{eqnarray}
    \tilde a_2 &=& -\frac{1}{4} \bigg\{2 \delta \mu  \sqrt{\frac{4 q (q+1) (zt)^2}{(2 \delta \mu +zt)^2}+1}\nonumber\\
    && +zt \pqty{\sqrt{\frac{4 q (q+1) (zt)^2}{(2 \delta \mu +zt)^2}+1}+2 q+1}-1\bigg\},
\end{eqnarray}
and setting $\tilde a_2=0$, one again finds the phase boundary which is in agreement with Eq.~\eqref{pb}.

Within this approximation, one can also obtain the integer filling line in the SF phase analytically. Namely, from $\prescript{}{i}{\bra{\Phi_0}} \hat n_i \ket{\Phi_0}_i = q+\cos\bar\chi \sin^2(\bar\theta/2) $, the integer filling condition means $\chi= \pi/2$. Thus Eq.~\eqref{chitheta} leads to $\bar\theta = \arccos\frac{-z\tilde t-4\delta \mu}{z\tilde t} $. One then solve $\bqty{\partial_\theta \varepsilon_\vari(\theta,\pi/2)}|_{\theta=\bar\theta}$ to get
\begin{equation}\label{phsmu}
    \delta \mu_{\phs} = -\frac{1}{4}\bqty{z \tilde t + \pqty{\sqrt{q+1}+ \sqrt{q}}^{-2}},
\end{equation}
which is bent down due to asymmetric Bose enhencement at finite fillings. It is this line we refer to as the PH symmetric line.

We define three commuting bosonic operators that create three Fock states at a given site $i$, $b_{i,\ell}^\dagger\ket{\vac} = \ket{q+\ell}_i$, $\ell=0,\pm1$. They must satisfy the local constraint $\sum_{\ell=-1}^1 b^\dagger_{i,\ell}b_{i,\ell}=1$. Then the original bosonic creation operator can be expressed as $\hat a_i^\dagger =\sum_{\ell,\ell'}\prescript{}{i}{\bra{q+\ell}}\hat a_i^\dagger\ket{q+\ell'}_i\hat b_{i,\ell}^\dagger \hat b_{i,\ell'}= \sqrt{q+1}\hat b_{i,1}^\dagger \hat b_{i,0} + \sqrt{q} \hat b_{i,0}^\dagger \hat b_{i,-1}$, and the strong-coupling mean-field Hamiltonian Eq.~\eqref{hmf1} becomes (up to a constant)
\begin{equation}\label{hmf2}
    \hat H_i^\mf = \begin{bmatrix}
        \hat b_{i,-1}^\dagger & \hat b_{i,0}^\dagger & \hat b_{i,1}^\dagger
    \end{bmatrix} H \begin{bmatrix}
        \hat b_{i,-1}\\
         \hat b_{i,0}\\
          \hat b_{i,1}\\
    \end{bmatrix}.
\end{equation}
where
\begin{equation}
    H = \begin{bmatrix}
        \frac{1}{2}+\delta\mu+zt\phi^2 & -\sqrt{q}zt \phi & 0\\
        -\sqrt{q} zt \phi & zt\phi^2 & -\sqrt{q+1} zt \phi\\
        0 & -\sqrt{q+1} zt \phi & \frac{1}{2}-\delta\mu+zt\phi^2
    \end{bmatrix}.
\end{equation}
Eq.~\eqref{hmf2} is diagonalized by the following rotation,
\begin{equation}
    \begin{bmatrix}
        \hat \beta_{i,0}\\
        \hat \beta_{i,1}\\
        \hat \beta_{i,2}
    \end{bmatrix} = T(\alpha) \begin{bmatrix}
        \hat b_{i,-1}\\
         \hat b_{i,0}\\
          \hat b_{i,1}\\
    \end{bmatrix},
\end{equation}
where
\begin{widetext}
\begin{equation}
    T(\alpha) = \left(
    \begin{array}{ccc}
     \sin \left(\frac{\bar{\theta }}{2}\right) \sin \left(\frac{\bar{\chi }}{2}\right) & \cos \left(\frac{\bar{\theta }}{2}\right) & \sin \left(\frac{\bar{\theta }}{2}\right) \cos \left(\frac{\bar{\chi }}{2}\right) \\
     \cos (\alpha ) \cos \left(\frac{\bar{\theta }}{2}\right) \sin \left(\frac{\bar{\chi }}{2}\right)+\sin (\alpha ) \cos \left(\frac{\bar{\chi }}{2}\right) & -\cos (\alpha ) \sin \left(\frac{\bar{\theta }}{2}\right) & \cos (\alpha ) \cos \left(\frac{\bar{\theta }}{2}\right) \cos \left(\frac{\bar{\chi }}{2}\right)-\sin (\alpha ) \sin \left(\frac{\bar{\chi }}{2}\right) \\
     \cos (\alpha ) \cos \left(\frac{\bar{\chi }}{2}\right)-\sin (\alpha ) \cos \left(\frac{\bar{\theta }}{2}\right) \sin \left(\frac{\bar{\chi }}{2}\right) & \sin (\alpha ) \sin \left(\frac{\bar{\theta }}{2}\right) & -\sin (\alpha ) \cos \left(\frac{\bar{\theta }}{2}\right) \cos \left(\frac{\bar{\chi }}{2}\right)-\cos (\alpha ) \sin \left(\frac{\bar{\chi }}{2}\right) \\
    \end{array}
    \right),
\end{equation}
\end{widetext}
with $\bar \chi = \chi(\bar \theta)$. The rotation angle $\alpha$ is determined by requiring that $THT^\dagger = \tmop{diag}(\varepsilon_0, \varepsilon_1, \varepsilon_2)$. It can be shown straightforwardly that $(THT^\dagger)_{12}=(THT^\dagger)_{13}=0$ always hold; one only needs to set $(THT^\dagger)_{23}=0$, which leads to
\begin{widetext}
\begin{equation}\label{alphabar}
    \bar \alpha = \frac{2 \cos \left(\frac{\bar{\theta }}{2}\right) \left(\sin \left(\bar{\chi }\right) \left(z t \cos \left(\bar{\theta }\right)+4 \delta \mu -z t\right)-2 \sqrt{q} \sqrt{q+1} z t \left(\cos \left(\bar{\theta }\right)-1\right) \cos \left(\bar{\chi }\right)\right)}{\cos \left(\bar{\theta }\right) \left(1-2 \delta \mu  \cos \left(\bar{\chi }\right)\right)+z t \sin ^2\left(\bar{\theta }\right) \left(2 \sqrt{q} \sqrt{q+1} \sin \left(\bar{\chi }\right)+2 q+1\right)+\cos \left(\bar{\chi }\right) \left(z t \sin ^2\left(\bar{\theta }\right)-6 \delta \mu \right)-1}.
\end{equation}
\end{widetext}
Note the rotated operators also satisfy the local constraint as before, and the local eigenstates are now given by $\ket{\Phi_\sigma}_i=\beta_{i,\sigma}^\dagger\ket{\vac}$, for $\sigma = 0,1,2$. Here $\sigma=0$ modes is the local ground state, and $\sigma=1,2$ modes are two local excitations.

In terms of the rotated operators $\hat\beta$, Eq.~\eqref{Ham} given in the main text can be recast as
\begin{eqnarray}\label{hsb}
    \hat H &=& -\sum_{ij}\sum_{\alpha\sigma\gamma\delta}\mathcal A_{\sigma\alpha} \mathcal A_{\gamma\delta}t_{ij}\beta_{i\alpha}^\dagger\beta_{i\sigma}\beta_{j\gamma}^\dagger\beta_{j\delta}\nonumber\\
    && +\sum_i\sum_{\alpha\sigma} \mel{\Phi_\alpha}{\bqty{\frac{1}{2}U \hat n_i (\hat n_i -1) - \mu  \hat n_i}}{\Phi_\sigma}\beta_{i\alpha}^\dagger \beta_{i\sigma},\nonumber\\
\end{eqnarray}
where $\mathcal A_{\alpha\sigma}= \prescript{}{i}{\bra{\Phi_\alpha}}\hat a_i^\dagger\ket{\Phi_\sigma}_i$. Note we have used the fact that $\mathcal A_{\alpha\sigma}$ is real, hence $\prescript{}{i}{\bra{\Phi_\alpha}}\hat a_i^\dagger\ket{\Phi_\sigma}_i=\prescript{}{i}{\bra{\Phi_\sigma}}\hat a_i\ket{\Phi_\alpha}_i$. By treating $\hat \beta$ operators as classical fields, the energy minimum is reached for $\beta_0^* = \beta_0 = 1$ with all other modes vanish. We perform a harmonic expansion of Eq.~\eqref{hsb} around this saddle point by condensing operators $\hat \beta_0^{(\dagger)}$:
\begin{equation}\label{replacement}
    \hat\beta_{i,0}^{(\dagger)} \rightarrow \sqrt{1-\sum_{\sigma>0}\hat\beta_{i,\sigma}^\dagger\hat\beta_{i,\sigma}}\approx 1-\frac{1}{2}\sum_{\sigma>0} \hat\beta_{i,\sigma}^\dagger \hat\beta_{i,\sigma}.
\end{equation}
The zeroth order term gives the mean-field ground-state energy $H^{(0)}= \expval{\hat H}{G}$. The first order term can be rearranged as $\hat H^{(1)} = \sum_i \sum_{\alpha>0} \hat\beta_{i\alpha}\prescript{}{i}{\bra{\Phi_0}}\hat H_i^\mf\ket{\Phi_\alpha}_i + \hc$, which vanishes identically due to orthogonality between local eigenstate $\ket{\Phi_0}_i$ and $\ket{\Phi_\sigma}_i$ for $\sigma>0$. The second order term reads
\begin{equation}\label{h2}
    \hat H^{(2)} = \hat H_\hopp +\hat H_\pair +\hat H_\onsite,
\end{equation}
where
\begin{subequations}\label{3h}
    \begin{align}
        \hat H_\hopp &= -\sum_{ij}\sum_{\alpha,\sigma>0} \pqty{\mathcal A_{0\alpha} \mathcal A_{0\sigma}t_{ij}+A_{\alpha 0} \mathcal A_{\sigma 0}t_{ji}} \hat\beta^\dagger_{i\alpha} \hat\beta_{j\sigma},\\
    \hat H_\pair &= -\frac{1}{2}\sum_{ij} \sum_{\alpha,\sigma>0} \pqty{\mathcal A_{0\alpha} \mathcal A_{\sigma0} t_{ij}+\mathcal A_{\alpha0}\mathcal A_{0\sigma} t_{ji}} \hat\beta_{i\alpha}^\dagger \hat\beta_{j\sigma}^\dagger +\hc,\\
    \hat H_\onsite &= \sum_i \sum_{\alpha,\sigma>0} \pqty{\varepsilon_\alpha - \varepsilon_0} \hat\beta_{i\alpha}^\dagger \hat\beta_{i\alpha},
    \end{align}
\end{subequations}
corresponding to hopping, pairing and on-site terms, respectively. Note, since $\mathcal A_{\alpha 0}$ is generally nonzero, the hopping and pairing terms will couple two local excitation modes. In momentum space, Eq.~\eqref{3h} becomes
\begin{subequations}\label{3hk}
    \begin{align}
        \hat H_\hopp & = \sum_{\vb k} \sum_{\alpha,\sigma>0} \sum_{\eta\eta'}  \bigg[\mathcal A_{0\alpha} \mathcal A_{0\sigma} (H_{\vb k}^\hop)_{\eta\eta'}\nonumber\\
        &\quad +\mathcal A_{\alpha0}\mathcal A_{\sigma0} (H_{-\vb k}^{\hop*})_{\eta\eta'} \bigg]\hat\beta^\dagger_{\vb k,\alpha\eta} \hat\beta_{\vb k,\sigma\eta'}, \\
        \hat H_\pair & = \frac{1}{2}\sum_{\vb k} \sum_{\alpha,\sigma>0} \sum_{\eta\eta'}  \bigg[\mathcal A_{0\alpha} \mathcal A_{\sigma0} (H^\hop_{\vb k})_{\eta\eta'}\nonumber\\
        &\quad +\mathcal A_{\alpha0}\mathcal A_{0\sigma} (H_{-\vb k}^{\hop*})_{\eta\eta'}\bigg]\hat \beta^\dagger_{\vb k,\alpha\eta} \hat \beta_{-\vb k,\sigma\eta'}^\dagger +\hc,\\
        \hat H_\onsite &= \sum_{\vb k} \sum_{\alpha>0} \sum_{\eta\eta'} (\varepsilon_{\alpha}-\varepsilon_0) \hat \beta_{\vb k,\alpha\eta}^\dagger \hat \beta_{\vb k,\alpha\eta'},
    \end{align}
\end{subequations}
where $\eta,\eta'$ are sublattice indices (assuming total $n_s$ sublattices), and we have assumed that the hopping matrix $t_{ij}$ is Hermitian, i.e., $t_{ji}=t_{ij}^*$. Here $H_{\vb k}^\hop$ is the Bloch Hamiltonian of the hopping term, specifically, for the two-dimensional SSH model with four sublattices, $n_s=4$, as shown in Fig.~\ref{fig1} of the main text, it is
\begin{equation}\label{2dsshhk}
    \resizebox{\hsize}{!}{$H_{\vb k}^\hop = -\left(
    \begin{array}{cccc}
     0 & t_1+t_2 \ee^{-\ii k_y} & t_1+t_2 \ee^{-\ii k_x} & 0 \\
     t_1+t_2 \ee^{\ii k_y} & 0 & 0 & t_1+t_2 \ee^{-\ii k_x} \\
     t_1+t_2 \ee^{\ii k_x} & 0 & 0 & t_1+t_2 \ee^{-\ii k_y} \\
     0 & t_1+t_2 \ee^{\ii k_x} & t_1+t_2 \ee^{\ii k_y} & 0 \\
    \end{array}
    \right).$}
\end{equation}
By arranging $\beta_{\vb k,\alpha\eta}$ into a vector $\bm\beta_{\vb k}$, and defining two $2n_s\times 2n_s$ matrices $A_{\vb k}$ and $B_{\vb k}$ with components
\begin{subequations}\label{abm}
    \begin{align}
        (A_{\vb k})_{\alpha\eta,\sigma\eta'} &= \mathcal A_{0\alpha} \mathcal A_{0\sigma} (H_{\vb k}^\hop)_{\eta\eta'} +\mathcal A_{\alpha0}\mathcal A_{\sigma0} (H_{-\vb k}^{\hop*})_{\eta\eta'}\nonumber\\
        &\quad +(\varepsilon_{\alpha}-\varepsilon_0)\delta_{\eta\eta'},\\
        (B_{\vb k})_{\alpha\eta,\sigma\eta'} &= \mathcal A_{0\alpha} \mathcal A_{\sigma0} (H_{\vb k}^\hop)_{\eta\eta'} +\mathcal A_{\alpha0}\mathcal A_{0\sigma} (H_{-\vb k}^{\hop*})_{\eta\eta'},
    \end{align}
\end{subequations}
Eq.~\eqref{h2} can be written compactly in a BdG form as
\begin{equation}\label{hbdgk}
    \hat H^{(2)} = \frac{1}{2} \sum_{\vb k} \begin{bmatrix}
        \bm\beta_{\vb k}^\dagger & \bm\beta_{-\vb k}
    \end{bmatrix} H_{\vb k}^\bdg \begin{bmatrix}
        \bm\beta_{\vb k}\\
        \bm\beta_{-\vb k}^\dagger
    \end{bmatrix},
\end{equation}
where
\begin{equation}\label{hbdgk1}
    H_{\vb k}^\bdg = \begin{bmatrix}
        A_{\vb k} & B_{\vb k}\\
        B_{-\vb k}^* & A_{-\vb k}^T
    \end{bmatrix}.
\end{equation}

For models with time reversal symmetry (TRS),
\begin{equation}\label{trs}
    H_{\vb k}^\hop = H_{-\vb k}^{\hop*},
\end{equation}
Eq.~\eqref{abm} simplifies to
\begin{subequations}\label{3hk1}
    \begin{align}
        A &= F \otimes H_{\vb k}^\hop + \mathcal E\otimes I_{n_s} \\
        B &= G \otimes H_{\vb k}^\hop
    \end{align}
\end{subequations}
where
\begin{subequations}\label{fge}
    \begin{align}
    F_{\alpha\sigma} &= \mathcal A_{0\alpha} \mathcal A_{0\sigma} +\mathcal A_{\alpha0}\mathcal A_{\sigma0}  \\
    G_{\alpha\sigma} &= \mathcal A_{0\alpha} \mathcal A_{\sigma0}+\mathcal A_{\alpha0}\mathcal A_{0\sigma} \\
    \mathcal E_{\alpha\sigma} &= \delta_{\alpha\sigma}(\varepsilon_\alpha- \varepsilon_0)
\end{align}
\end{subequations}
are two-by-two, real symmetric matrices in local excitation space, and $I_{n_s}$ is the identity matrix in sublattice space.

Under OBCs, order parameter becomes site-dependent, and the mean-field theory has to be worked out numerically in a self-consistent manner from Eq.~\eqref{hmf}. Then the physical boson annihilation operator in the rotated local basis $\mathcal A_i$, and local eigenenergies $\varepsilon_{i\alpha}$, $\alpha=1,2,3$, become site-dependent. Taking into account these differences, the quadratic Hamiltonian Eq.~\eqref{h2} can still be solved by a Bogoliubov transformation. In our numerics, for simplicity, we reuse the order parameter obtained from PBCs. This leads to a tiny gap near zero energy for the excitations, which is manually removed by a shift of chemical potential as in Ref.~\cite{xu2016pi}. We have numerically checked that this gap tends to zero as we enlarge the system size. Moreover, topological properties of the highly excited states are not affected anyway.

\section{Diagonalization of the BdG Hamiltonian}
In this section, we first review the process of diagonalization of a generic bosonic BdG Hamiltonian, which also severs to introduce notations and set the stage for the discussion of following sections. Then we show that in the large filling limit, this Bogoliubov transformation can be constructed analytically. In particular, we explicitly show that the coupling between phase modes and amplitude modes only vanish when the noninteracting Hamiltonian has TRS, even in the large filling limit. We only discuss the momentum space version, the real space version can be formulated similarly.

\subsection{The general case}
A generic bosonic BdG Hamiltonian, as given in Eq.~\eqref{hbdgk1}, is diagonalized by a Bogoliubov transformation,
\begin{equation}
    W_{\vb k}^\dagger H_{\vb k}^\bdg W_{\vb k} = D_{\vb k} = \tau_0 \otimes \begin{bmatrix}
        E_{1,\vb k} & & & \\
        & E_{2,\vb k} &\\
        & & \ddots \\
    \end{bmatrix},
\end{equation}
where from here on $\tau_0$ and $\tau_{1,2,3}$ denote 2-by-2 identity matrix and Pauli matrices acting on the Nambu space, and
\begin{equation}\label{wmat}
    W_{\vb k} = \begin{bmatrix}
        U_{\vb k} & V_{-\vb k}^*\\
        V_{\vb k} & U_{-\vb k}^*
    \end{bmatrix}
\end{equation}
is a pseudo-unitary matrix satisfying
\begin{equation}
    W_{\vb k}^\dagger \Sigma_3 W_{\vb k} = \Sigma_3 \qand W_{\vb k} \Sigma_3 W_{\vb k}^\dagger = \Sigma_3.
\end{equation}
We define the Bogoliubov quasi-particle annihilation operator via $\va*{\bm\beta}_{\vb k} = W_{\vb k}\va*{\bm\gamma}_{\vb k} $, where $\va*{\bm\beta}_{\vb k} = (\bm\beta_{\vb k}, \bm\beta_{-\vb k}^\dagger)^T$. More explicitly,
\begin{equation}\label{betagamma}
    \bm\beta_{\vb k} = U_{\vb k}\bm\gamma_{\vb k} + V_{-\vb k}^* \bm\gamma_{-\vb k}^\dagger \qand \bm\beta_{\vb k}^\dagger = V_{-\vb k} \bm\gamma_{-\vb k} + U_{\vb k}^* \bm\gamma_{\vb k}^\dagger.
\end{equation}
Then Eq.~\eqref{hbdgk} after this Bogoliubov transformation becomes
\begin{eqnarray}
    \hat H^{(2)} &=& \frac{1}{2} \sum_{\vb k} \va*{\bm\beta}_{\vb k}^\dagger H_{\vb k}^\bdg \va*{\bm\beta}_{\vb k}\nonumber\\
    & =& \frac{1}{2}\sum_{\vb k} \va*{\bm\gamma}_{\vb k}^\dagger W^\dagger_{\vb k} H_{\vb k}^\bdg W_{\vb k} \va*{\bm\gamma}_{\vb k}\nonumber\\
    & =& \sum_{\vb k,\lambda} (E_{\lambda} + \frac{1}{2})\gamma_{\vb k\lambda}^\dagger \gamma_{\vb k\lambda} 
\end{eqnarray}
where $\lambda$ is the band index. Generally this Bogoliubov transformation has to be done numerically.

\subsection{Analytical solution at the large filling limit}
Firstly, we note that in the large filling limit, $q\gg1$, the original Hamiltonian Eq.~(1) in the main text when truncated to three local states, Eq.~\eqref{trunc3}, can be mapped to a bond-staggered XY model with uniaxial single-ion anisotropy and magnetic coupling \cite{altman2002oscillating}, $\hat H \sim \sum_{i,j}\tilde t_{ij}(\hat S_i^x \hat S_j^x + \hat S_i^y \hat S_j^y) + \sum_i \bqty{\frac{1}{2}U(\hat S_i^z)^2 -\delta \mu\hat S_i^z}$, where $\langle i,j \rangle$ denotes nearest neighbors and $\tilde t_{ij} = pt_{ij}$. In the following, we will focus on the PH symmetric line, i.e., $\delta \mu =0$. When taking $q\gg1$, Eq.~\eqref{chitheta} becomes $\chi=\pi/2$, and Eq.~\eqref{gse} becomes $\varepsilon_\vari = \frac{1}{2}\pqty{\sin^2\frac{\theta}{2}-z\tilde t\sin^2\theta}$. Its minimization leads to $\bar\theta = \arccos\frac{U}{4qzt}$ if $U<4qzt$, and $\bar\theta = 0$ otherwise. Then the rotation angle $\bar\alpha$ given in Eq.~\eqref{alphabar} reduces to $\bar\alpha = 0$. Thus the unitary matrix $T$ is
\begin{equation}
    T(\bar\theta ) = \frac{1}{\sqrt{ 2}} \begin{bmatrix}
         \sin \frac{\bar{\theta }}{2} & \sqrt{2} \cos \frac{\bar{\theta }}{2} & \sin
           \frac{\bar{\theta }}{2} \\
         \cos \frac{\bar{\theta }}{2} & -\sqrt{2}\sin \frac{\bar{\theta }}{2} & \cos
           \frac{\bar{\theta }}{2} \\
         1 & 0 & -1
    \end{bmatrix}
\end{equation}
And the physical boson annihilation operator in this rotated basis is
\begin{equation}\label{alargeq}
    \mathcal A_i \approx T\begin{bmatrix}
         0 & \sqrt{q} & 0 \\
         0 & 0 & \sqrt{q} \\
         0 & 0 & 0 
    \end{bmatrix} T^\dagger = \sqrt\frac{q}{2} \begin{bmatrix}
     \sin \bar\theta & \cos \bar\theta & -\cos \frac{\bar\theta }{2} \\
     \cos \bar\theta & -\sin \bar\theta & \sin \frac{\bar\theta }{2} \\
     \cos \frac{\bar\theta }{2} & -\sin \frac{\bar\theta }{2} & 0
    \end{bmatrix}.
\end{equation}
Then $\hat H_\hopp$ and $\hat H_\pair$ defined in Eq.~\eqref{3hk} becomes
\begin{widetext}
\begin{subequations}\label{hhplargeq}
        \begin{align}
            \hat H_\hopp & = \frac{q}{2} \sum_{\vb k} \bm\beta^\dagger_{\vb k} \begin{bmatrix}
                (H_{\vb k}^\hop+H_{-\vb k}^{\hop*}) \cos^2\theta & (-H_{\vb k}^\hop+H_{-\vb k}^{\hop*}) \cos\frac{\theta}{2} \cos\theta\\
                (-H_{\vb k}^\hop+H_{-\vb k}^{\hop*}) \cos\frac{\theta}{2}  & \frac{1}{2}(H_{\vb k}^\hop+H_{-\vb k}^{\hop*}) \cos^2\frac{\theta }{2}
            \end{bmatrix} \bm\beta_{\vb k}, \\
            \hat H_\pair & = \frac{q}{4}\sum_{\vb k} \bm\beta^\dagger_{\vb k} \begin{bmatrix}
                (H_{\vb k}^\hop+H_{-\vb k}^{\hop*}) \cos^2\theta & (H_{\vb k}^\hop-H_{-\vb k}^{\hop*}) \cos\frac{\theta}{2} \cos\theta\\
                (H_{\vb k}^\hop - H_{-\vb k}^{\hop*}) \cos\frac{\theta}{2}  & -\frac{1}{2}(H_{\vb k}^\hop+H_{-\vb k}^{\hop*}) \cos^2\frac{\theta }{2}
            \end{bmatrix} \bm\beta_{-\vb k}^\dagger +\hc.
        \end{align}
\end{subequations}
\end{widetext}
Importantly, if and only if the system has TRS, i.e., when Eq.~\eqref{trs} holds, two local excitation modes become \emph{decoupled}. In other words, three matrices defined in Eq.~\eqref{fge} all become diagonal,
\begin{subequations}
    \begin{align}
        F &= q \begin{bmatrix}
            \cos^2\bar\theta & 0\\
            0 & \cos^2\frac{\theta}{2}
        \end{bmatrix},\\
        G &= q\begin{bmatrix}
            \cos^2\theta & 0\\
            0 & -\cos^2\frac{\theta}{2}
        \end{bmatrix},\\
        \mathcal E &= \begin{bmatrix}
            2z\tilde t\sin^2\bar\theta +\frac{1}{2}U\cos\bar\theta & 0\\
            0 & z\tilde t\sin^2\bar\theta + \frac{1}{2}U \cos^2\frac{\bar\theta}{2}
        \end{bmatrix}.
    \end{align}
\end{subequations}
Thus the second order term, Eq.~\eqref{h2}, can be written as
\begin{eqnarray}\label{h2largeq}
    \hat H^{(2)} &=& \frac{1}{2}\sum_{\vb k} \begin{bmatrix}
        \bm\beta_{\vb k,A}^\dagger & \bm\beta_{-\vb k,A} & \bm\beta_{\vb k,P}^\dagger & \bm\beta_{-\vb k,P}
    \end{bmatrix}\nonumber\\
    &&\times \begin{bmatrix}
        H_{\vb k,A}^\bdg & \\
        & H_{\vb k,P}^\bdg
    \end{bmatrix} \begin{bmatrix}
        \bm\beta_{\vb k,A}\\
        \bm\beta_{-\vb k,A}^\dagger\\
        \bm\beta_{\vb k,P}\\
        \bm\beta_{-\vb k,P}^\dagger
    \end{bmatrix}
\end{eqnarray}
where
\begin{eqnarray}\label{hbdgap}
    H_{\vb k,\alpha}^\bdg &=& \begin{bmatrix}
        \xi_{\alpha} + \kappa_{\alpha} H_{\vb k}^\hop & \zeta_{\alpha} \kappa_{\alpha}H_{\vb k}^\hop\\
        \zeta_\alpha \kappa_{\alpha} H_{\vb k}^\hop & \xi_\alpha +\kappa_\alpha H_{\vb k}^\hop
    \end{bmatrix}\nonumber\\
    & =& \xi_\alpha\tau_0\otimes I_{n_s} + \kappa_\alpha\tau_0\otimes H_{\vb k}^\hop + \zeta_{\alpha}\kappa_{\alpha}\tau_1 \otimes H_{\vb k}^\hop ,\nonumber\\
\end{eqnarray}
with all parameters given in Table~\ref{tab1} of the main text. Note we have renamed the local excitation mode $\alpha = 1 \rightarrow A$ and $\alpha = 2 \rightarrow P$, which represents the amplitude modes and phase modes, respectively. This nomenclature will be justified in the next section.

A remarkable property of Eq.~\eqref{hbdgap} is that one can construct its Bogoliubov transformation analytically \cite{kumar2020dirac}. One first performs a unitary rotation using $\tilde Q_{\vb k} = \tau_0 \otimes Q_{\vb k}$, where $Q_{\vb k}$ diagonalizes the Bloch Hamiltonian $H_{\vb k}^\hop$, $Q_{\vb k}^\dagger H_{\vb k}^\hop Q_{\vb k} = D_{\vb k}$, with $D_{\vb k} = \tmop{diag}(d_1,\cdots, d_{n_s})$. It then leads to
\begin{equation}\label{hbdgktilde}
    \tilde Q_{\vb k}^\dagger H_{\vb k,\alpha}^\bdg \tilde Q_{\vb k} = \xi_\alpha \tau_0\otimes I_{n_s} + \eta_\alpha \tau_0 \otimes D_{\vb k} + \zeta_\alpha\eta_{\alpha} \tau_1 \otimes D_{\vb k}.
\end{equation}
One then performs another pseudo-unitary transformation using
\begin{eqnarray}\label{pmatrix}
    P_{\vb k,\alpha} &=& \tau_0 \otimes \begin{bmatrix}
        \cosh \beta_{1,\vb k,\alpha} & &\\
        & \ddots & \\
        && \cosh \beta_{n_s,\vb k,\alpha}
    \end{bmatrix}\nonumber\\
    && + \tau_1 \otimes \begin{bmatrix}
        \sinh \beta_{1,\vb k,\alpha} & &\\
        & \ddots & \\
        && \sinh \beta_{n_s,\vb k,\alpha}
    \end{bmatrix},
\end{eqnarray}
where $\cosh\beta_{i,\vb k,\alpha} = \sqrt{\frac{\xi_\alpha + d_{i,\vb k}}{2 E_{i,\vb k,\alpha}}+\frac{1}{2}}$, $\sinh\beta_{i,\vb k,\alpha} = -\tmop{sign}(d_{i,\vb k}) \sqrt{\frac{\xi_\alpha+d_{i,\vb k}}{2E_{i,\vb k,\alpha}}-\frac{1}{2}}$ and
\begin{equation}\label{eigenenergy}
    E_{i,\vb k,\alpha} (d_{i\vb k}) = \sqrt{\xi_\alpha^2+2\xi_\alpha d_{i,\vb k}}.
\end{equation}
It then fully diagonalizes Eq.~\eqref{hbdgktilde},
\begin{equation}
    P_{\vb k,\alpha}^\dagger \tilde Q_{\vb k}^\dagger H_{\vb k,\alpha}^\bdg \tilde Q_{\vb k} P_{\vb k,\alpha} = \tau_0 \otimes \begin{bmatrix}
        E_{1,\vb k,\alpha} & & & \\
        & E_{2,\vb k,\alpha} & & \\
        & & \ddots & \\
        & & & E_{n_s,\vb k,\alpha}
    \end{bmatrix}.
\end{equation}
Thus the pseudo-unitary matrix generally defined in Eq.~\eqref{wmat} now becomes
\begin{equation}
    W_{\vb k,\alpha} = \tilde Q_{\vb k} P_{\vb k,\alpha},
\end{equation}
or, more explicitly,
\begin{subequations}\label{uvlargeq}
    \begin{align}
        U_{\vb k,\alpha} &= Q_{\vb k}C_{\vb k,\alpha},\\
        V_{\vb k,\alpha} &= Q_{\vb k} S_{\vb k,\alpha},
    \end{align}
\end{subequations}
where
\begin{subequations}
	\begin{align}
		C_{\vb k} &= \begin{bmatrix}
            \cosh \beta_{1,\vb k,\alpha} & &\\
            & \ddots & \\
            && \cosh \beta_{n_s,\vb k,\alpha}
        \end{bmatrix},\\
        S_{\vb k} &= \begin{bmatrix}
            \sinh \beta_{1,\vb k,\alpha} & &\\
            & \ddots & \\
            && \sinh \beta_{n_s,\vb k,\alpha}
        \end{bmatrix}.
	\end{align}
\end{subequations}
\section{Phase-amplitude character}
Here we discuss how to determine the phase-amplitude character of these excitation modes. Particular, we show explicitly that the large filling limit of SSH-BHM has two types of excitations with pure phase and pure amplitude character, respectively. We only consider the momentum space version, the real space version can be formulated similarly.

Consider the oscillation of the order parameter induced by a small perturbation above the ground state. For a perturbation characterized by an excitation labeled by momentum $\vb k$ and band index $\lambda$, the perturbed state evolves in time as $\ket{\Psi_{\vb k,\lambda}(t)} = \ee^{-\ii \hat H t} \pqty{\ket{G} + \epsilon \gamma^\dagger_{\vb k\lambda}\ket{G} } $, with $\epsilon\ll 1$. Thus the oscillation of the order parameter around the ground state expectation value, $\delta\phi_i(t) = \expval{\hat a_{i}}{\Psi_{\vb k,\lambda}(t)} - \expval{\hat a_i}{G}$, to linear order in $\epsilon$, reads
\begin{align}
   & \delta\phi_i(t)\nonumber\\
    &\propto \mel{G}{\hat a_{i} \ee^{-\ii \hat H^{(2)} t}\hat \gamma_{\vb k\lambda}^\dagger}{G} + \mel{G}{\hat\gamma_{\vb k\lambda} \ee^{\ii \hat H^{(2)} t} \hat a_i}{G}\\
    &= \mel{G}{\hat a_{i} \hat \gamma_{\vb k\lambda}^\dagger}{G} \ee^{-\ii \omega_{\vb k\lambda} t} + \mel{G}{\hat\gamma_{\vb k\lambda}  \hat a_i}{G} \ee^{\ii \omega_{\vb k\lambda} t}\\
    &= \sum_{\alpha}\bigg[\mathcal A_{\alpha0}\pqty{ \mel{G}{\hat \beta_{i\alpha}^\dagger  \hat \gamma_{\vb k\lambda}^\dagger}{G}\ee^{-\ii \omega_{\vb k\lambda} t}+ \mel{G}{\hat\gamma_{\vb k\lambda}\hat \beta_{i\alpha}^\dagger }{G} \ee^{\ii \omega_{\vb k\lambda} t}  }\nonumber \\
    &\quad +\mathcal A_{0\alpha}\pqty{ \mel{G}{\hat \beta_{i\alpha} \hat \gamma_{\vb k\lambda}^\dagger}{G}\ee^{-\ii \omega_{\vb k\lambda} t} + \mel{G}{\hat\gamma_{\vb k\lambda}\hat \beta_{i\alpha}}{G} \ee^{\ii \omega_{\vb k\lambda} t} }\bigg]\\
    &= \sum_{\vb p,\alpha}\bigg\{\mathcal A_{\alpha0}\bigg[ \mel{G}{\hat \beta_{\vb p\alpha\eta}^\dagger  \hat \gamma_{\vb k\lambda}^\dagger}{G}\ee^{-\ii (\omega_{\vb k\lambda} t+\vb p\cdot\vb r_i)}\nonumber\\
    &\quad + \mel{G}{\hat\gamma_{\vb k\lambda}\hat \beta_{\vb p\alpha\eta}^\dagger }{G} \ee^{\ii (\omega_{\vb k\lambda} t-\vb p\cdot\vb r_i)}  \bigg]\nonumber \\
    &\quad +\mathcal A_{0\alpha}\bigg[ \mel{G}{\hat \beta_{\vb p\alpha\eta} \hat \gamma_{\vb k\lambda}^\dagger}{G}\ee^{-\ii (\omega_{\vb k\lambda} t-\vb p\cdot\vb r_i)}\nonumber\\
    &\quad + \mel{G}{\hat\gamma_{\vb k\lambda}\hat \beta_{\vb p\alpha\eta}}{G} \ee^{\ii (\omega_{\vb k\lambda} t + \vb p\cdot\vb r_i) }\bigg]\bigg\}
\end{align}
Using Eq.~\eqref{betagamma} and the fact that $\gamma_{\vb k\lambda}\ket{G}=0$, one has
\begin{equation}
    \delta\phi_i(t) \propto X_{\vb k}\ee^{-\ii (\omega_{\vb k\lambda} t-\vb k\cdot\vb r_i)} + Y_{\vb k} \ee^{\ii (\omega_{\vb k\lambda} t-\vb k\cdot\vb r_i)}
\end{equation}
where
\begin{subequations}\label{xydef}
    \begin{align}
        X_{\vb k} &= \sum_{\alpha}\bqty{ \mathcal A_{\alpha0} (V_{\vb k})_{\alpha\eta,\lambda} +\mathcal A_{0\alpha} (U_{\vb k})_{\alpha\eta,\lambda}}\\
        Y_{\vb k} &= \sum_{\alpha}\bqty{ \mathcal A_{\alpha0}(U_{\vb k}^*)_{\alpha\eta,\lambda} +\mathcal A_{0\alpha} (V_{\vb k}^*)_{\alpha\eta,\lambda}}.
    \end{align}
\end{subequations}
Here the row of matrices $U$ and $V$ is labeled by two indices, the local excitation $\alpha$ and sublattice $\eta$.
Thus the imaginary (real) part of the order parameter oscillation is
\begin{subequations}\label{ripart}
        \begin{align}
        \tmop{Re} \delta \phi &\propto X_{\vb k} + Y_{\vb k}\\
        \tmop{Im} \delta \phi &\propto X_{\vb k} - Y_{\vb k}
    \end{align}
\end{subequations}
A pure amplitude (phase) oscillation of the order parameter corresponds to $\tmop{Im} \delta \phi = 0$ ($\tmop{Re} \delta \phi = 0$), we define a flatness parameter
\begin{equation}\label{flatnessdef}
    F=\frac{\abs{\tmop{Re} \delta\phi} - \abs{\tmop{Im} \delta\phi}}{\abs{\tmop{Re} \delta\phi} + \abs{\tmop{Im} \delta\phi}} \in [-1,1] ,
\end{equation}
to quantify the amplitude and phase components of an excitation: A positive (negative) flatness indicates dominant amplitude (phase) character. A pure amplitude (phase) oscillation corresponds to $F=1 $ $(-1)$.

In the large filling limit, the flatness defined in Eq.~\eqref{flatnessdef} can be obtained analytically. Using Eq.~\eqref{alargeq} and Eq.~\eqref{uvlargeq}, Eq.~\eqref{xydef} becomes, for $\alpha = A$ mode
\begin{subequations}
    \begin{align}
        X_{\vb k,A} &=  \pqty{Q_{\eta\lambda}\sinh \beta_{\lambda,\vb k,A} + Q_{\eta\lambda}\cosh\beta_{\lambda,\vb k,A} }\cos\bar\theta,\\
        Y_{\vb k,A} &=  \pqty{ Q_{\eta\lambda} \cosh \beta_{\lambda,\vb k,A}+Q_{\eta\lambda} \sinh\beta_{\lambda,\vb k,A} }\cos\bar\theta,
    \end{align}
\end{subequations}
hence their difference vanishes. While for $\alpha=P$ mode
\begin{subequations}
    \begin{align}
        X_{\vb k,P} &=  \pqty{\sinh \beta_{\lambda,\vb k,P} Q_{\eta\lambda} -  \cosh\beta_{\lambda,\vb k,P} Q_{\eta\lambda}}\cos\frac{\bar\theta}{2},\\
        Y_{\vb k,P} &= \pqty{\cosh \beta_{\lambda,\vb k,P} Q_{\eta\lambda} - \sinh\beta_{\lambda,\vb k,P} Q_{\eta\lambda}}\cos\frac{\bar\theta}{2},
    \end{align}
\end{subequations}
hence their sum vanishes. Here we have fixed the gauge by requiring $Q_{\eta\lambda}$ to be real. It then follows from Eq.~\eqref{ripart} that for $A$ mode $ \delta \phi $ is purely real, while for $P$ mode $ \delta \phi $ is purely imaginary, and the flatness is $+1$ and $-1$ for the $A$ mode and $P$ mode, respectively, which justifies the nomenclature.

\section{Topological character}
In this section, we discuss how to define the band topology for the bosonic BdG system of the 2D SSH-BHM. In particular, we prove that the topological index, namely, the symplectic polarization vector, is quantized to a $\mathbb Z_2$ number due to the inversion symmetry. And relate it to the parity eigenvalues at the inversion symmetric momenta. Then we show explicitly that in the infinite filling limit, the amplitude and phase bands have the same topological index as the underlying noninteraction Hamiltonian.

For a bosonic BdG system with the inversion symmetry defined by
\begin{equation}\label{isbdg}
    \mathcal I_\tau  H_{\vb k}^\bdg \mathcal I_\tau ^{-1} = H_{-\vb k}^\bdg,
\end{equation}
the symplectic polarization defined in 1D by Engelhardt and Brandes \cite{engelhardt2015topological} can be straightforwardly generalized to the vectorized version,
\begin{equation}
    \vb P = \frac{1}{(2\pi)^2} \int_\bz \dd[2]{k} \vb A(\vb k),
\end{equation}
where the symplectic $\mathrm{U(1)}$ Berry connection is $ A_\mu(\vb k) = \ii \sum_{\lambda_1\leq \lambda\leq \lambda_2} \tmop{Tr}\pqty{\Gamma_\lambda W_{\vb k}^{-1} \partial_\mu W_{\vb k}}$. We define a sewing matrix $B_{\vb k} = W_{-\vb k}^\dagger \Sigma_3 \mathcal I_\tau  W_{\vb k}$, with $\Sigma_3 = \tau_3\otimes I$ being the Pauli spin-$z$ matrix acting on the Nambu space, to relate eigenstates at momenta $\vb k$ and its inversion symmetric partner at $-\vb k$. Note this sewing matrix is pseudo-unitary, block diagonal and satisfies $B_{-\vb k}^\dagger = B_{\vb k} $. Due to the inversion symmetry, we can relate the symplectic Berry connection at $\vb k$ to $-\vb k$, by using this sewing matrix
\begin{equation}
    A_\mu(-\vb k)= - A_\mu(\vb k) + \ii \partial_{\mu}\ln \det B^<_{\vb k},
\end{equation}
where $B_{\vb k}^<$ denotes the projection of $B_{\vb k}$ to the block consisting of bands between $\lambda_1\leq \lambda\leq \lambda_2$. Then for $\mu=1$ (and similarly for $\mu=2$), we have
\begin{align}
  P_1 &= \frac{1}{(2\pi)^2} \int_{-\pi}^\pi \dd{k_2} \int_0^\pi \dd{k_1} [A_1(k_1,k_2) + A_1(-k_1,k_2)] \nonumber \\
  &= \frac{1}{(2\pi)^2} \int_{-\pi}^\pi \dd{k_2} \int_0^\pi \dd{k_1} [A_1(k_1,k_2)\nonumber\\
 &\quad - A_1(k_1,-k_2) + \ii \partial_{k_1}\ln \det B^<_{\vb k}] \nonumber \\
  &= \frac{1}{2\pi} \int_{-\pi}^\pi \dd{k_2}\bqty{\frac{\ii}{2\pi} \int_0^\pi \dd{k_1} \partial_{k_1}\ln (\det B^<_{\vb k})}.
\end{align}
Notice the integral over $k_1$ gives the winding number of $\det B_{\vb k}^<$, a pure phase at a fixed $k_2$. Since the system under consideration has TRS, namely, the Chern number always vanishes, which means that one can always find a continuous gauge. Thus the winding number can not change discontinuously along $k_2$ direction, and we can simply evaluate this constant by taking $k_2=0$, which leads to
\begin{equation}
    P_1 = \frac{\ii}{2\pi}\int_0^\pi\dd{k_1}\partial_{k_1} \ln \det B_{k_1,0}^< = \frac{\ii}{2\pi} \ln\frac{\det B_{\vb X_1}^<}{\det B_{\vb \Gamma}^<}.
\end{equation}
Since at the inversion symmetric momenta, we have $\Sigma_3 B_{\vb k_\inv} = \Sigma_3 W_{\vb k_\inv}^\dagger \Sigma_3 \mathcal I_\tau  W_{\vb k_\inv}= W_{\vb k_\inv}^{-1} \mathcal I_\tau W_{\vb k_\inv}$, which leads to $\det B_{\vb k_\inv}^< = \pm \prod_{\lambda_1\leq \lambda \leq \lambda_2} \eta_{\vb k_\inv} $ (plus/minus sign for particle/hole space), where $\eta$ is the eigenvalue of the inversion operator $\mathcal I_\tau $. In conlusion, each component of $\vb P$ is restrictedly quantized to a $\mathbb Z_2$ number,
\begin{equation}
    P_\mu = \frac{1}{2} \pqty{ \sum_{\lambda_1\leq \lambda \leq \lambda_2} n_{\lambda,\mu} \mod 2 },
\end{equation}
where $(-1)^{n_{\lambda,\mu}} = \eta_\lambda(X_\mu)\eta_\lambda(\Gamma)$.

At the infinite filling limit, since two modes are decoupled, we can study Eq.~\eqref{hbdgap} for $\alpha=A,P$, individually. This BdG Hamiltonian is easily seen to satisfy the inversion symmetry Eq.~\eqref{isbdg} with
\begin{equation}\label{iobdg}
    \mathcal I_\tau  = \tau_0 \otimes \mathcal I.
\end{equation}
Since $E_{i,\vb k,\alpha} (d_{i\vb k})$ defined in Eq.~\eqref{eigenenergy} is a monotonically increasing function, the band gap closes at the same $t_1/t_2$ and at the same $\vb k$ points in the BZ for the noninteracting bands $d_{i\vb k}$ and the BdG bands $E_{i,\vb k,\alpha}$. Moreover, the parity eigenvalues for these two systems are also the same:
\begin{align}
    W^{-1}_{\vb k} \mathcal I_\tau  W_{\vb k} & = \Sigma_3 W^\dagger_{\vb k} \Sigma_3 (\tau_0 \otimes \mathcal I) W_{\vb k}\nonumber \\
    & = \Sigma_3 P^\dagger_{\vb k} \tilde Q_{\vb k}^\dagger \Sigma_3 (\tau_0\otimes \mathcal I) \tilde Q_{\vb k} P_{\vb k}\nonumber \\
    & = \Sigma_3 P^\dagger_{\vb k} [\tau_3\otimes (Q_{\vb k}^{-1} \mathcal I Q_{\vb k} )] P_{\vb k}\nonumber \\
    & = \tau_0 \otimes (Q_{\vb k}^{-1} \mathcal I Q_{\vb k} ),
\end{align}
where to arrive at the last line, we have used the explicit form of matrix $P_{\vb k}$ given in Eq.~\eqref{pmatrix}. We thus conclude that the topological phase boundary does not alter.

In the finite filling case, our bosonic BdG Hamiltonian still enjoys the inversion symmetry $\tilde I = I_4\otimes \mathcal I $. We numerically find that the topological transition point again occurs at $t_1=t_2$, which is understandable from the symmetric roles played by two hopping parameters.

\section{A Ginzburg-Landau analysis via strong-coupling random phase approximation}
Here we derive the effective action used in the main text, following a strong-coupling random phase approximation developed by Sengupta and Dupuis \cite{sengupta2005mott}, and discuss the condition leads to the (approximate) PHS.

In the imaginary-time path integral formalism, the Euclidean action is
\begin{eqnarray}
    S &=& \int_0^\beta \dd{\tau} \bigg[\sum_ia^*_i \partial_\tau a_i - \sum_{ij} t_{ij}a^*_i a_j -\mu\sum_i a_i^* a_i\nonumber\\
    && + \frac{1}{2}U\sum_i \abs{a_i}^4\bigg].
\end{eqnarray}
Using a Hubbard-Stratonovich (HS) transformation, we introduce an auxiliary field $\varphi$ to decouple the hopping term and integrate out the original field $a$, the partition function formally becomes
\begin{equation}\label{hs1}
    Z = Z_0 \int \mathcal D[\varphi^*,\varphi] \ee^{-\int_0^\beta\dd\tau\sum_i \varphi^*_i(\tau)(t^{-1})_{ij} \varphi_j(\tau) + W_\loc[\varphi^*,\varphi]}.
\end{equation}
The generating functional for connected $l$-particle local Green's function is defined by
\begin{align}
    &G_\loc^{c,(l)}(\tau_1,\dots,\tau_l;\tau_1',\dots,\tau_l')\nonumber \\
     &= (-1)^l \langle a_i(\tau_1) \cdots a_i(\tau_l) a_i^*(\tau_l') \cdots a_i^*(\tau_1')\rangle_\loc \nonumber \\
    &= (-1)^l \frac{\delta^{(2l)}W_\loc[\varphi^*,\varphi]}{\delta\varphi^*_i(\tau_1)\cdots \delta\varphi_i^*(\tau_l)\delta\varphi_i (\tau'_l) \cdots \delta\varphi_i(\tau'_1)}.
\end{align}
Here the local Hamiltonian is
\begin{equation}\label{loch}
    \hat H_\loc = \sum_i\bqty{\frac{1}{2}U \hat n_i (\hat n_i -1) - \mu  \hat n_i},
\end{equation}
and $\expval{\cdots}_\loc$ means that the average is taken with respect to $\hat H_\loc$. Upon reverting above equation, we obtain
\begin{eqnarray}
    &&W_\loc [\varphi^*,\varphi] = \sum_{l=1}^\infty \frac{(-1)^l}{(l!)^2}\int \dd\tau_1 \cdots \dd\tau_l\nonumber\\
    &&G_\loc^{c,(l)} (\tau_1,\dots,\tau_l;\tau_1',\dots,\tau_l') \varphi^*(\tau_1) \cdots\varphi^*(\tau_l)\varphi(\tau_l')\cdots\varphi(\tau_1').\nonumber\\
\end{eqnarray}
The effective action given in Eq.~\eqref{hs1} is used by Ref.~\cite{fisher1989boson,sachdev2011quantum} to study the quantum phase transition between the SF phase and the MI phase. However, as pointed out by Sengupta and Dupuis \cite{sengupta2005mott}, in the SF phase, the Green's function obtained is not physical, thus the excitation spectrum is out of reach. More importantly, it is hard to investigate the topology associated with the inverse of the hopping matrix, being generally a complicated infinite-range hopping matrix.

We can kill two birds with one stone by performing a second HS transformation. This process decouples the hopping term in Eq.~\eqref{hs1}, where the pure local $\varphi$ field can be integrated out again. Since the correlation function built from the auxiliary field introduced in this second HS transformation and the original bosonic fields $a$ are the same (the proof is easy and can be found in Ref.~\cite{sengupta2005mott}), we use the same notation for them. The resulting effective action is
\begin{widetext}
\begin{align}
    S^\eff &= \int\dd \tau\dd\tau' \sum_{ij} \bigg\{a^*_i(\tau) \bigg[-\Gamma_\loc^{(1)} (\tau;\tau') \delta_{ij}+t_{ij} \delta(\tau-\tau')\bigg]a_j(\tau')\bigg\}\nonumber \\
    &+ \int\dd\tau_1\dd\tau_2 \dd\tau_1'\dd\tau_2' \sum_i \bqty{\frac{1}{4}\Gamma_\loc^{(2)}(\tau_1,\tau_2;\tau_2',\tau_1')a^*_i(\tau_1)a^*(\tau_2)a_i(\tau_2')a_i(\tau'_1) } + \cdots ,
\end{align}
\end{widetext}
where $\cdots$ denotes higher-order local vertex functions, which are neglected. Here, the one-particle local vertex function is given by $\Gamma_\loc^{(1)} = [G_\loc^{(1)}]^{-1}$ (from here on we will omit the superscript in $G^{(1)}$ for the single-particle Green's function). While the two-particle local vertex function $\Gamma_\loc^{(2)}$ can be obtained from the one- and two-particle connected local Green's function using the standard formula \cite{negele2018quantum},
\begin{eqnarray}
    &&G_\loc^{c,(2)} (\tau_1,\tau_2;\tau_1',\tau_2') = -\int_0^\beta \dd \tau_3\dd\tau_4\dd\tau_3'\dd\tau_4'G_\loc (\tau_1;\tau_3)\nonumber\\
    && G_\loc(\tau_2;\tau_4)\Gamma_\loc^{(2)}(\tau_3,\tau_4;\tau_3',\tau_4')G_\loc(\tau_3';\tau_1')G_\loc (\tau_4';\tau_2').\nonumber\\
\end{eqnarray}

Before proceeding further, we review the local problem defined by Eq.~\eqref{loch}. For a given site, it is already diagonal in particle number basis. The ground state has $q$ bosons with $q=\lfloor \mu/U\rfloor+1$ if $\mu>0$, and $q=0$ otherwise, and the corresponding energy $e_q = -\mu q+(U/2)q(q-1)$. The single-particle Green's function is (for $\tau>0$)
\begin{eqnarray}
    G_\loc(\tau;0) &=& -\expval{T_\tau a(\tau)a^\dagger(0)}\nonumber\\
    & =& -\frac{1}{Z_\loc} \sum_{n=0}^\infty (n+1)\ee^{-(\beta-\tau)e_n-\tau e_{n+1}},
\end{eqnarray}
where $Z_\loc = \sum_{n=0}^\infty \ee^{-\beta e_n}$. In Matsubara frequency space at zero temperature, it becomes
\begin{eqnarray}\label{g1f}
    G_\loc (\ii \omega )|_{T=0} &=& \lim_{\beta\rightarrow\infty} \int_0^\beta \dd\tau \, G_\loc(\tau;0)\ee^{\ii\omega\tau}\nonumber\\
     &=& \frac{-q}{\ii\omega+e_{q-1} -e_q}+\frac{q+1}{\ii \omega+e_q -e_{q+1}}.
\end{eqnarray}
The two-particle Green's function can be obtained similarly, whose explicit expression in the static limit at zero temperature, $\bar G_\loc^{c,(2)}$, can be found in Ref.~\cite{sengupta2005mott}. If we approximate $\Gamma_\loc^{(2)}$ by its static value, $\bar\Gamma_\loc^{(2)}=-\bar G_\loc^{c,(2)}/\bar G_\loc^4$, and introducing $\tilde U = \frac{1}{2}\bar \Gamma_\loc^{(2)}$, the effective action then reads
\begin{widetext}
\begin{equation}
    S^\eff = \int\dd\tau\dd\tau' \sum_{ij} a^*_i(\tau) \bqty{-G_\loc^{-1}(\tau;\tau')\delta_{ij} +t_{ij}\delta(\tau-\tau') }a_j(\tau')     +\frac{1}{2}\tilde U\int\dd\tau\sum_i \abs{a_i(\tau)}^4.
\end{equation}
\end{widetext}
In Matsubara frequency space, using Eq.~\eqref{g1f}, one can expand $G_\loc^{-1}(\ii \omega)$ around $\omega= 0$:
\begin{eqnarray}\label{gtoc}
    -G_\loc^{-1}(\ii\omega) &=& \frac{(-qU+\mu+\ii\omega)(U-qU+\mu+\ii\omega)}{U+\mu+\ii\omega}\nonumber\\
      &=& \sum_{\ell=0}^\infty c_\ell (\ii\omega)^\ell,
\end{eqnarray}
where $c_\ell = (\ell!)^{-1} \partial^l G_\loc^{-1}(\ii\omega)/\partial(\ii\omega)^l |_{\omega=0} $, and the most important coefficient is
\begin{equation}
    c_1 = -1 + \frac{q(1+q)U^2}{(U+\mu)^2},
\end{equation}
whose vanishing on the $\mu-t$ phase diagram is the so-called particle-hole (PH) symmetric line $\mu_\phs = \bqty{\sqrt{q(1+q)}-1}U$. Note it starts at the tip of the $q$th lobe and is a horizontal line independent of $zt$, which overlooks the hopping effects in comparison with Eq.~\eqref{phsmu}. Precisely at the PH symmetric line, Eq.~\eqref{gtoc} becomes
\begin{eqnarray}
    &&- G^{-1}_\loc(\ii\omega)|_{\mu=\mu_\phs}\nonumber\\
    & =& \frac{\bqty{-1+\ii \omega -q+\sqrt{q(1+q)}}\bqty{\ii\omega-q+\sqrt{q(1+q)}}}{\ii\omega+\sqrt{q(1+q)}},\nonumber\\
\end{eqnarray}
which leads to
\begin{equation}
    c_\ell|_{\mu=\mu_\phs} = \bqty{\frac{-1}{\sqrt{q(1+q)}}}^{\ell-1} =  \mathcal O(q^{-\ell+1}),\qfor \ell > 1.
\end{equation}
Thus, at the large filling limit and on the PH symmetric line, only $c_2$ survives even away from the low-energy limit.

\section{A brief discussion on validity of slave boson method}
Generally speaking, as a strong-coupling expansion, our approach is expected to work well in the Mott-insulating phase and in the superfluid phase close to the SF-MI phase transition boundary; and become worse in the weakly interacting limit (where the standard Bogoliubov theory should be more appropriate).

More specifically, we note that the local Hilbert space is enlarged when the slave bosons are introduced at each site; however, this redundancy is then removed by imposing the local constraint. There are two main approximations involved: (1) only three local states at each sites are considered. (2) the local constraint is actually broken when condensing $\beta_G$ and making the replacement Eq.~\eqref{replacement}, with higher-order terms omitted.

Regarding to the first issue, we note that, in the vicinity of the Mott phase, number fluctuations are small, which allows one to truncate the Hilbert space into the subspace of the lowest local number states. This local number fluctuations have also been experimentally measured \cite{greiner2002quantum}, and found indeed to be suppressed due to strong interactions near the vicinity of the Mott phase. Moreover, this approximation can be systematically improved by the inclusion of further local states. The error occurred by this truncation can be computed by comparing the two cases. We have numerically checked that such error is indeed small for parameter regions of our interests. In fact, by including these extra local states, we find that the resulting spurious excitations are almost equal two-,three-,... particle excitations of the mean-field Hamiltonian Eq.~(A1), reflecting the fact that they are high-energy excitations, outside of our low-energy theory in the strong-coupling limit.

Regarding to the second issue, the same approximation also occurs in the widely used method of Holstein-Primakoff boson \cite{auerbach2012interacting} and the standard Bogoliubov theory \cite{kawaguchi2012spinor}. One way to verify the validity of this approximation is to check that, \textit{a posteriori}, the quantum depletion $\sum_{\alpha\neq G}\expval{\beta_{i,\alpha}^\dagger \beta_{i,\alpha}}$ is indeed quite small comparing to unit. A similar calculation has been performed in \cite{huber2007dynamical} (for the standard Bose-Hubbard model in 2D): this quantity is around $0.2$ and is largest at the phase transition point. Therefore, the expansion Eq.~\eqref{replacement} is justified and it should be a good approximation for the parameter region of our interests. Interaction among the Higgs and the Goldstone modes can be studied in the future by including these higher order terms in the expansion Eq.~\eqref{replacement}.

%


\end{document}